\documentclass{article}

\usepackage[final]{neurips_2025}

\usepackage[utf8]{inputenc} %
\usepackage[T1]{fontenc}    %
\usepackage{graphicx}
\usepackage{subcaption}
\usepackage{ulem}
\usepackage{amsmath}
\usepackage{mathtools}
\usepackage{multirow}
\usepackage{hyperref}       %
\usepackage{wrapfig}
\setcitestyle{round}
\hypersetup{
   colorlinks=true,
   linkcolor=[RGB]{30, 30, 180},
   citecolor=[RGB]{30, 30, 180},
   urlcolor=magenta,
   pdfborder=0 0 0,
   pdftitle={},
   pdfsubject={}, 
   pdfkeywords={},
   pdfauthor={},%
   pdfstartview=FitH
}
\usepackage[capitalize,noabbrev]{cleveref}
\usepackage[nolist,nohyperlinks]{acronym}
\usepackage{pifont}
\usepackage{booktabs}
\usepackage[table]{xcolor} %
\usepackage[inline]{enumitem}

\usepackage{array}
\usepackage{amsmath}
\usepackage{amssymb}
\usepackage{mathtools}
\usepackage{amsthm}
\usepackage{bm}

\newcommand{\ourmethod}{GyroSwin}

\newcommand{\ourmethodlinear}{$\text{GyroSwin}_{\text{Linear}}$}
\newcommand{\ourmethodsmall}{$\text{GyroSwin}_{\text{Small}}$}
\newcommand{\ourmethodmedium}{$\text{GyroSwin}_{\text{Medium}}$}
\newcommand{\ourmethodlarge}{$\text{GyroSwin}_{\text{Large}}$}

\newcommand\Bb{\bm{b}}

\newcommand\Bq{\bm{q}}
\newcommand\Br{\bm{r}}

\newcommand\Bv{\bm{v}}

\newcommand\Bx{\bm{x}}

\newcommand\BA{\bm{A}}
\newcommand\BB{\bm{B}}

\newcommand\BE{\bm{E}}

\newcommand\BK{\bm{K}}

\newcommand\BQ{\bm{Q}}
\newcommand\BR{\bm{R}}

\newcommand\BV{\bm{V}}
\newcommand\BW{\bm{W}}
\newcommand\BX{\bm{X}}

\makeatletter
\newcommand{\dlmf}[1]{%
\citep[%
  \def\nextitem{\def\nextitem{, }}%
  \@for \el:=#1\do{\nextitem\href{http://dlmf.nist.gov/\el}{(\el)}}%
]{olver_nist_2010}%
}
\makeatother

\newcommand{\textbfp}[1]{\vspace{0.5em}\noindent\textbf{#1.}}

\newcommand{\xmark}{\cellcolor{red!20}\ding{55}}   %

\title{
  \raisebox{-1.75pt}{\includegraphics[width=15pt]{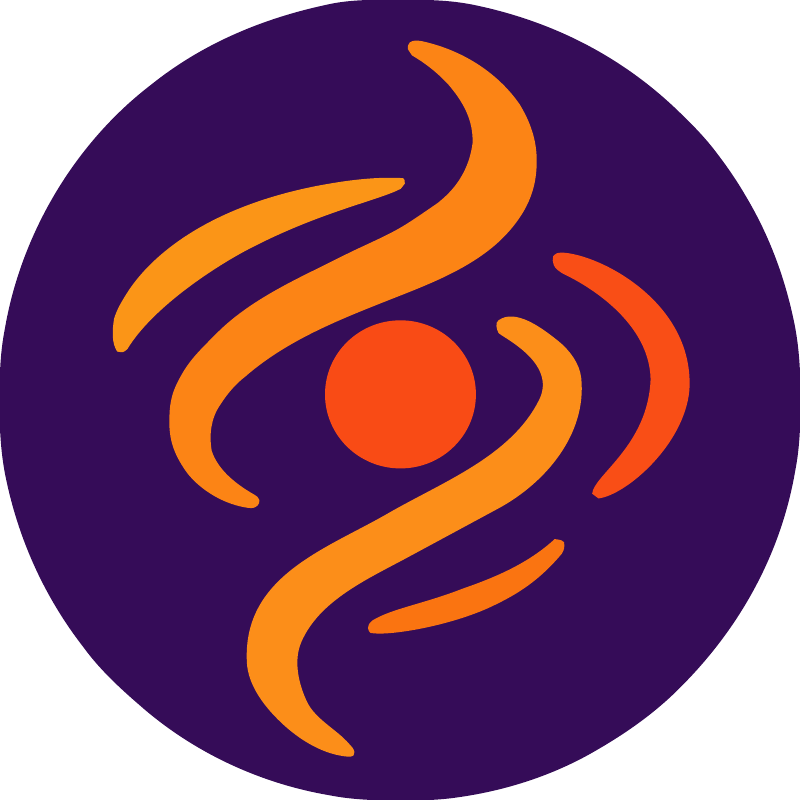}}\hspace{4pt}%
  \ourmethod{}: 5D Surrogates for Gyrokinetic \\ Plasma Turbulence Simulations
}

\author{
  \textbf{Fabian Paischer} \thanks{Equal contribution} $\:\:^{1,3}$ \hspace{16px}%
  \textbf{Gianluca Galletti} \footnotemark[1] $\:\:^{1}$ \hspace{16px}%
  \textbf{William Hornsby}$^{2}$ \hspace{16px}%
  \textbf{Paul Setinek}$^1$ \vspace{6px}\\%
  \textbf{Lorenzo Zanisi}$^{2}$ \hspace{16px}%
  \textbf{Naomi Carey}$^{2}$ \hspace{16px}%
  \textbf{Stanislas Pamela}$^2$ \hspace{16px}%
  \textbf{Johannes Brandstetter}$^{1,3}$
  \vspace{8px}\\
  {$^1$~ELLIS Unit, Institute for Machine Learning, JKU Linz}\\
  {$^2$~United Kingdom Atomic Energy Authority, Culham campus}\\
  {$^3$~EMMI AI, Linz}
  \\\vspace{6px}
  \texttt{\{paischer,galletti,brandstetter\}@ml.jku.at}
  \\\vspace{3px}
  \href{https://github.com/ml-jku/neural-gyrokinetics}{\raisebox{-0.3em}{\includegraphics[width=1.15em]{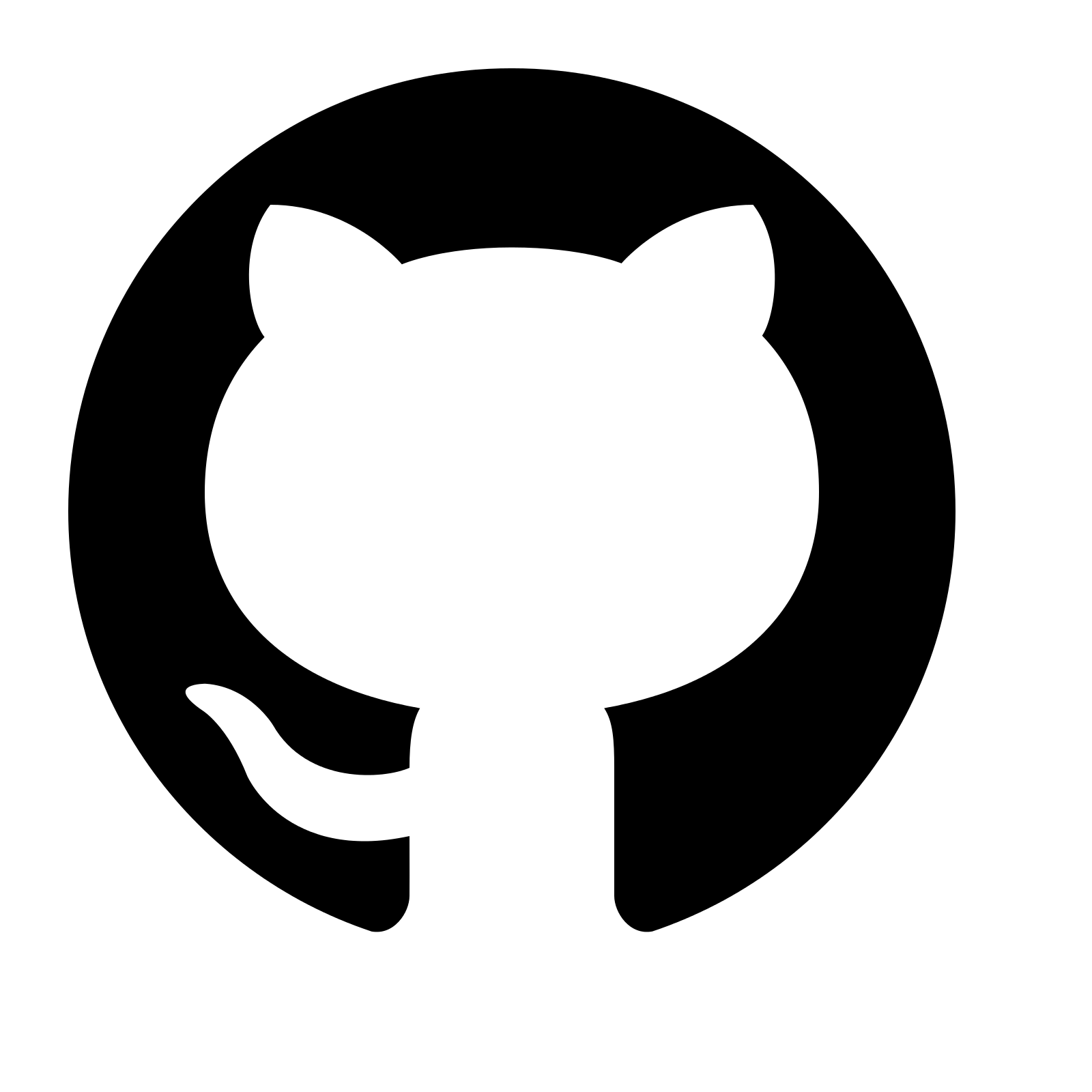}} \texttt{ml-jku/neural-gyrokinetics}},\hspace{1em}
\href{https://huggingface.co/papers/2510.07314}{\raisebox{-0.3em}{\includegraphics[width=1.15em]{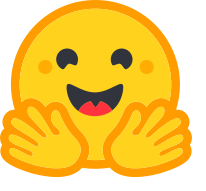}} \texttt{ml-jku/gyroswin}}
\vspace{-12px}
}

\begin{document}

\maketitle

\begin{abstract}
    Nuclear fusion plays a pivotal role in the quest for reliable and sustainable energy production.
    A major roadblock to viable fusion power is understanding plasma turbulence, which significantly impairs plasma confinement, and is vital
    for next-generation reactor design. 
    Plasma turbulence is governed by the nonlinear gyrokinetic equation, which evolves a 5D distribution function over time.
    Due to its high computational cost, reduced-order models are often employed in practice to approximate turbulent transport of energy.
    However, they omit nonlinear effects unique to the full 5D dynamics.
    To tackle this, we introduce \ourmethod{}, the first scalable 5D neural surrogate that can model 5D nonlinear gyrokinetic simulations, thereby capturing the physical phenomena neglected by reduced models, while providing accurate estimates of turbulent heat transport.
    \ourmethod{}
    \begin{enumerate*}[label=(\roman*)]
        \item extends hierarchical Vision Transformers to 5D, 
        \item introduces cross-attention and integration modules for latent 3D$\leftrightarrow$5D interactions between electrostatic potential fields and the distribution function, and
        \item performs channelwise mode separation inspired by nonlinear physics.
    \end{enumerate*}
    We demonstrate that \ourmethod{} outperforms widely used reduced numerics on heat flux prediction, captures the turbulent energy cascade, and reduces the cost of fully resolved nonlinear gyrokinetics by three orders of magnitude while remaining physically verifiable.
    \ourmethod{} shows promising scaling laws, tested up to one billion parameters, paving the way for scalable neural surrogates for gyrokinetic simulations of plasma turbulence. 
\end{abstract}

\section{Introduction}
\label{sec:intro}
Nuclear fusion promises sustainable energy by fusing hydrogen isotopes within an ionized plasma.
Because the plasma reaches temperatures of hundreds of millions of degrees, it must be confined by a magnetic field.
During confinement, turbulence can arise from micro-instabilities, leading to energy and particle transport toward the reactor walls.
This causes plasma to leak from its magnetic cage, resulting in heat and density loss, and therefore impaired energy production.
The design and control of plasma scenarios strictly require knowledge of turbulent transport, which can be obtained via nonlinear gyrokinetic simulations that evolve a 5D Partial Differential Equation (PDE) over time.

The computational cost of nonlinear gyrokinetic simulations is prohibitive.
Therefore, QuasiLinear approaches (QL), such as QuaLiKiz \citep{Bourdelle2015,Citrin2017} and TGLF \citep{Staebler2007,Staebler2010}, are commonly used to approximate turbulent transport.
They are based on a Reduced-Order Model (ROM) that operates in 3D and adopt a so-called saturation rule to estimate nonlinear fluxes.
These saturation rules usually employ free parameters that are fitted to nonlinear flux data.
However, QL models neglect essential parts of the nonlinear physics that contribute substantially to evolving turbulent transport, namely zonal flows.
Therefore, a reliable estimate of turbulent fluxes to date is only attainable via expensive nonlinear gyrokinetic simulations.

To tackle the prohibitive cost of nonlinear gyrokinetic simualtions, we introduce \ourmethod{}, a scalable neural surrogate model to efficiently approximate turbulent transport.
\ourmethod{} is based on three essential ingredients:
\begin{enumerate*}[label=(\roman*)]
    \item extension of a Swin transformer \citep{liu_swin_2021} to 5D data,
    \item ilatent cross-attention and integration modules for interaction between 5D and 3D fields, as well as 5D $\rightarrow$ 3D integration, and
    \item channelwise mode separation informed by nonlinear physics.
\end{enumerate*}
To ground \ourmethod{} on physically meaningful quantities, we train it in a multitask manner on the 5D distribution function and derived quantities thereof, such as 3D potential fields, and scalar fluxes.

\ourmethod{} accurately captures nonlinear physics of 5D gyrokinetics.
To verify this, we infer physical downstream quantities of the predicted 5D PDE, such as heat flux, turbulence related spectra, and zonal flows.
We find that the predicted quantities align well with the ground truth for unseen simulations and \ourmethod{} scales favorably compared to other neural surrogate approaches.
Furthermore, \ourmethod{} is three orders of magnitude faster than the numerical code GKW for the adiabatic electron approximation.
In summary, our contributions are as follows.
\begin{itemize}[itemsep=1pt, topsep=0.5pt, leftmargin=*]
\item We introduce \ourmethod{}, a multitask hierarchical Vision Transformer that handles data of arbitrary dimensionality. It is trained on the adiabatic electron approximation using channelwise mode separation to predict the 5D PDE, 3D electrostatic potentials, and scalar fluxes simultaneously.
\item We propose latent cross-attention between fields of varying dimensionality, as well as integration layers that enable integrating along various dimensions of the 5D field.
\item We demonstrate that \ourmethod{} captures nonlinear dynamics in the 5D field, improves time-averaged flux predictions over state-of-the-art ROMs, and significantly outperforms deep learning surrogates on plasma turbulence modelling.
\item We demonstrate promising scaling laws of \ourmethod{}, tested up to 1B parameters.
\end{itemize}

\begin{figure}[t]
    \centering
    \includegraphics[width=\textwidth]{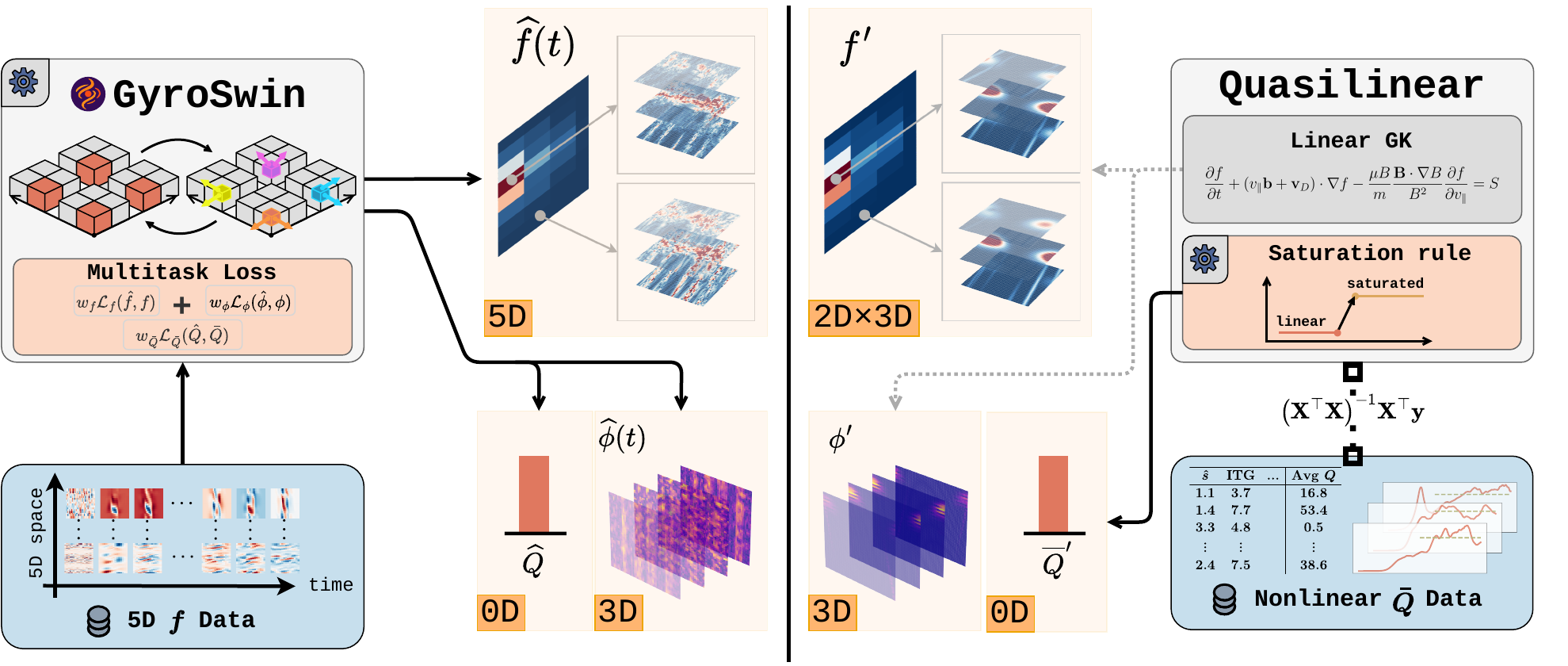}
    \caption{\textbf{Left:} \ourmethod{} models the 5D distribution function of nonlinear gyrokinetics and incorporates integration blocks to predict 3D electrostatic potential fields and scalar heat flux. \textbf{Right:} ROMs (quasilinear) solve a cartesian product of 2D modes in spectral space and 3D fields. Furthermore, they rely on saturation rules to approximate the nonlinear flux spectrum.
    }
    \label{fig:fig1}
\end{figure}

\section{Background and Related Work}
\label{sec:backgrnd}

\textbfp{5D Gyrokinetics} 
To determine turbulent transport, one must simulate the evolution of electrons and ions in the plasma over time.
In principle, this could be achieved by modelling each species as a particle in the plasma \citep{Birdsall_PIC_2005,Tskhakaya_PIC_2008}.
However, because the number of particles in a plasma can exceed 10$^{20}$, this approach is computationally expensive.
An alternative approach is to statistically model ensembles of particles in the plasma using a distribution function.
This approach is termed plasma kinetics and is desirable to analyse stability in a bulk of plasma.

Plasma kinetics models the time evolution of electrons and ions
via the distribution function $f$ based on 3D coordinates, their parallel and perpendicular velocities, and the angle around the field lines.
Gyrokinetics \citep{Krommes_gyrokinetics_2012} is a reduced form of plasma kinetics that is more efficient by averaging out the angle around the field lines and only considering a particle's guiding center.
Local gyrokinetics focuses on perpendicular spatial scales comparable to the gyroradius and on frequencies much lower than the particle cyclotron frequency, the circular motion of charged particles around magnetic field lines due to the Lorentz force.
Hence, the 5D gyrokinetic distribution function can be written as $f = f(k_{x},k_{y}, s, v_{||}, \mu)$, where
$k_x$ and $k_y$ are spectral coordinates (for the spatial $x$ and $y$),
$s$ is the toroidal coordinate along a field line, and $v_{||}$ and $\mu$ represent parallel and perpendicular velocity components, respectively.
The perturbed time-evolution of $f$, for each species (ions and electrons), is governed by 
\begin{equation}
\underbrace{\colorbox{blue!10}{$\displaystyle
\frac{\partial f}{\partial t} + (v_\parallel {\Bb} + {\Bv}_D) \cdot \nabla f   
-\frac{\mu B}{m} \frac{\BB \cdot \nabla B}{B^2} \frac{\partial f}{\partial v_\parallel}$}}_\text{Linear}
\; + \;
\underbrace{\colorbox{red!30}{$\displaystyle
\smash{{\Bv}_\chi \cdot \nabla f}
\vphantom{\frac{\partial f}{\partial v_\parallel}}$}}_\text{Nonlinear} 
 = S \ , 
\label{gyrovlas}
\end{equation}
where $v_\parallel \Bb$ is the motion along magnetic field lines, $\Bb = \BB / B$ is the unit vector along the magnetic field $\BB$ with magnitude $B$\footnote{We adopt uppercase notation for vector fields $\BE$ and $\BB$ to adhere with literature.}, $\Bv_D$ the magnetic drift due to gradients and curvature in $\BB$, and $\Bv_\chi$ describes
drifts arising from the $\BE \times \BB$ force, a key driver of plasma turbulence.
The nonlinear term models the interaction between the distribution function $f$ and the electrostatic potential, $-\nabla \phi=\BE$, which comes from the  velocity space integral of $f$ itself, and describes turbulent advection. The resulting nonlinear coupling constitutes the computationally most expensive term. Finally, $S$ is the source term that represents collisions between particles. For a  more detailed derivation of the gyrokinetic equation, see \cref{app:gyrokinetics}.

\textbfp{Quantities of interest}
In practice, the main quantity of interest is the radial transport of energy (heat) towards the reactor wall, which is crucial for reactor design.
It can be obtained by integrals on the 5D distribution function as 
\begin{equation}
\label{eq:flux_int}
Q = \int \mathbf{C} \int \mathbf{v}^2 \bm{\phi} \bm{f} \: \mathrm{d}v_{\parallel}\mathrm{d}\mu \:\: \mathrm{d}x\mathrm{d}y\mathrm{d}s, \qquad
\bm{\phi} = \mathbf{A} \int \mathbf{J_{0}} \bm{f} \: \mathrm{d}v_{\parallel}\mathrm{d}\mu, \quad
\end{equation}
where $\mathbf{A}, \mathbf{C} \in \mathbb{R}^{x\times s \times y}$ comprise geometric and operating parameters, 
$\mathbf{v} \in \mathbb{R}^{v_{\parallel} \times \mu}$ denotes particle energy,
and $\mathbf{J_{0}}$ denotes the zeroth-order Bessel function and $\bm{\phi} \in \mathbb{R}^{x \times s \times y} $ is the electrostatic potential.
In a tokamak, $\BB$ largely points in the toroidal direction, $k_x$ encodes the radial direction, and $k_y$ is perpendicular to $k_x$ (binormal $\approx$ poloidal) in Fourier space.
Electrostatic fluctuations that drive turbulence occur mainly in the $k_y$ direction, while $k_x$ encodes the radial structural properties of turbulence, i.e. the size of emerging eddies.

Radial turbulent transport occurs only if the amplitude of the $k_y$ modes is non-zero, as this results in $\BE \times \BB$ drift, and consequently in fully developed nonlinear turbulence.
A turbulent simulation usually follows a certain pattern.
In the linear phase, different modes in $k_y$ start growing due to underlying micro-instability, resulting in an initial increase in heat flux.
Afterwards, the simulations enter the nonlinear (saturated) regime where modes start interacting and the system converges into a statistically steady state.
This state is controlled by zonal flows that shear and break up arising eddies, effectively dampening turbulence \citep{Itoh_ZF_06}.

To analyse turbulence, practitioners usually investigate various spectra along the $k_y$ direction, as those provide insights into turbulent transport.
Most importantly, $Q(k_y)$ provides insight into which modes dominate the turbulence energy:
\begin{equation}
\label{eq:q_spectrum}
Q(k_y) = \sum_{v_{\parallel}, \mu, s, k_x} \bm{Q}(v_{\parallel}. \mu, s, k_x, k_y)  \:,
\end{equation}
where $\bm{Q}\in \mathbb{R}^{v_{\parallel} \times \mu \times s \times k_x \times k_y}$ is the flux field, which aggregates to the scalar flux $Q$ in \cref{eq:flux_int}.
In addition $W(k_y)\in\mathbb{R}^{k_y}$ describes the intensity of turbulence along $y$, and zonal flows dampen turbulence resulting in the statistically steady state which can be isolated as the first mode of $\bm{\phi}$ in $k_y$ direction: 
\begin{equation}
W(k_y) = \sum_{s, k_x} |\bm{\phi}(k_x, s, k_y)|^2 \:,\quad
\bm{\phi}_{\mathrm{ZF}}(x,t)
= \sum_{k_x} \phi_{k_x,\, k_y = 0}(t) \, e^{i k_x x}.
\label{eq:ky_zf}
\end{equation}
Since the system converges to a statistically steady state, practitioners usually investigate time-averaged quantities of $Q(k_y)$, $W(k_y)$, $\bm{\phi}_{\mathrm{ZF}}$, and scalar heat flux trace $\bar{Q}$.

\textbfp{State-of-the-art numerical approximations}
Solving \cref{gyrovlas} numerically is computationally expensive, especially for high-fidelity simulations across ion and electron scales.
Quasilinear (QL) models mitigate the main source of computational cost and are commonly adopted in integrated modelling pipelines \citep{VanMulders2021,citrin2024toraxfastdifferentiabletokamak}.
They are based on linear simulations that neglect the nonlinear term in \cref{gyrovlas} and solve for multiple $f$ for each mode in $k_x\times k_y$ independently.
This results in a speedup as the order of $f$ is reduced to $\hat{f} = f(v_{||}, \mu, s)$.
As a result, linear simulations do not account for interaction between modes, but only compute linear growth rates and potential transport contributions for each of them.
This mode-wise transport contribution is combined via so-called \textit{saturation rules} by
\begin{align}\label{saturation_rule}
    \bar{Q}= \int Q_{\text{QL}}(k_y) \hat{W}(k_y) dk_y \ ,
\end{align}
where $Q_{\text{QL}}(k_y)$ is the flux contribution per mode, and $\hat{W}(k_y)$ is a modelled weighting function which approximates $W(k_y)$ and whose free parameters are fit to nonlinear simulations.
Since QL models are based on linear simulations, they neglect nonlinear physics that substantially impact turbulence.
Another limitation of QL models are operating parameter regions leading to strong turbulence \citep{Staebler_quasilinear_2024,Kiefer_quasilinear_2021,dimits_shift_2000,Bourdelle_validity_2008}.

\textbfp{Machine learning for Gyrokinetics}
Machine Learning offers a fruitful alternative to ROMs.
Most literature to date has focused on multilayer perceptrons  \citep[MLP]{plassche_qlknn_2020,Citrin_qlknn_2023,Zanisi2024} or gaussian processes \citep{Hornsby2024}.
They usually map from operating parameters, such as temperature or density gradients to $\bar{Q}$ predicted via ROMs. 
Hence, they are inherently limited by the capabilities of ROMs used to produce the training data.
Few works have attempted to train machine learning surrogates for nonlinear gyrokinetics.
\citet{narita_toward_2022} leverage convolutional neural networks on 2D wavenumber ($k_x \times k_y$) slices to predict time to saturation and heat flux.
Building on this idea, \citet{mitsuru_multimodal_2023} include snapshots of electrostatic potentials as additional channels.
More recently, \citet{Wan_itgtransfer_2025} investigate transfer from low-fidelity to higher-fidelity simulations in a reduced 1D space.
Our proposed \ourmethod{} is fundamentally different, as it is trained directly on the 5D distribution function of nonlinear gyrokinetics.

\textbfp{Neural surrogates}
Over recent years, deep neural network-based surrogates have emerged as a computationally efficient alternative in science and engineering \citep{thuerey_physics_2021,zhang_artificial_2023,brunton_machine_2020}, impacting weather forecasting \citep{thorsten_fourcastnet_2023,bi_accurate_2023,lam_learning_2023,nguyen_climax_2023,bodnar_aurora_2024}, protein folding \citep{jumper_highly_2021,abramson_accurate_2024}, material
design \citep{merchant_scaling_2023,zeni_generative_2025,yang_mattersim_2024}, and multi-physics modelling~\cite{alkin2024neuraldem}. 
These success stories share the common thread of deep learning surrogates overcoming seemingly insurmountable challenges~\citep{brandstetter2024envisioning}. 
Especially for weather modelling, Vision Transformer~\citep[ViT]{dosovitskiy_image_2021} and especially Swin Transformer~\citep{liu_swin_2021} have shown exceptional performance, which manifested in the first medium-range weather modelling surpassing numerical accuracy \citep{bi_accurate_2023}, and the current state-of-the-art foundation model of the atmosphere\citep{bodnar_aurora_2024}.
In gyrokinetics we face similar challenges in terms of fidelity, complexity and local couplings.
However, these challenges are even exacerbated for gyrokinetics due to its 5D nature.

\section{\ourmethod{}}
\label{sec:method}

State-of-the-art ROMs neglect 5D physical phenomena that are crucial for reliable turbulent transport estimates.
To remedy this, we develop a neural surrogate, \ourmethod{}, that directly learns to evolve the 5D distribution function of nonlinear gyrokinetics over time.
The most important aspect in designing such a surrogate model is scalability as nonlinear gyrokinetic equations can span multiple turbulence modalities for different species at varying resolutions. 
Furthermore, downstream physical quantities, such as electrostatic potential fields and scalar fluxes should be consistent with the predicted 5D field.
Finally, understanding the structure of the 5D field enables us to bake inductive biases into the model architecture that improve capturing nonlinear dynamics, i.e., zonal flows.
Based on these observations we pose the following desiderata for designing 
\ourmethod{}.
\begin{enumerate}[leftmargin=*]
    \item \textbf{Scalability.} We identify two main candidates:
    \begin{enumerate*}[label=(\arabic*)]
        \item Convolution-based \citep[CNN]{fukushima_neocognitron_1980,lecun_backpropagation_1989} or FNOs \citep{li2021fourierneuraloperatorparametric}, and 
        \item Vision Transformers \citep[ViT]{dosovitskiy_image_2021}.
    \end{enumerate*}
    To preserve locality of the 5D field, CNNs or FNOs require factorized implementations \citep{wang2017factorized, tran2023factorizedfourierneuraloperators} that do not scale well (see \cref{tab:main_diagnostics}).
    ViTs are in principle applicable to 5D, however flattening a 5D field results in extremely long patch sequences.
    Due to their quadratic complexity \citep{vaswani_attention_2017}, ViTs do not scale well to high resolution input. 
    Hierarchical processing with linear attention, as in Swin \citep{liu_swin_2021}, provides an efficient way to preserve locality via local window attention and was successfully applied to 3D \citep{liu_video_2022} or 4D \citep{kim2023swift} input.
    Finally, patch embeddings can be compressed to a lower resolution to facilitate scalability. 
    \item \textbf{Modelling latent integrals.} 
    Electrostatic potentials and heat fluxes are computed as integrals (\cref{eq:flux_int}). We replicate this operation in the latent space of \ourmethod{}.
    To obtain the 3D latent fields, we introduce a latent integrator module that aggregates over the velocity space.
    Similarly, taking inspiration from \cref{eq:flux_int}, the flux prediction is based on cross-attention pooling of 3D and 5D latents.
    \item \textbf{Inductive biases.}
    The main benefit of \ourmethod{} over ROMs is that nonlinear dynamics can be captured that severely affect emerging turbulence.
    To bias \ourmethod{} towards nonlinear zonal flows, we perform \textit{channelwise mode separation}.
    Specifically, we separate the zonal flow mode from \cref{eq:ky_zf} from the other modes and transfer the spectral space to real space to retain the same dimensions.
    Then we add real and imaginary parts of the zonal flow as additional channel.
\end{enumerate}

Taking into account these requirements, we design \ourmethod{} as a Swin-based UNet \citep{ronneberger_unet_2015} with multiple branches to accommodate multitask predictions, and physical priors.
The following paragraphs explain the main architectural components of \ourmethod{}. We describe them in the domain-specific case of gyrokinetics with 5D shifted Window Attention (5DWA) and 5D up/downsampling layers, initially introduced by \cite{galletti20255dneuralsurrogatesnonlinear}.
All of them are generalized to n-dimensions through adaptive window partitioning and local convolution.

\begin{figure}[t]
    \centering
    \begin{subfigure}{.4\textwidth}
        \centering
        \raisebox{0px}{\includegraphics[width=0.8\textwidth]{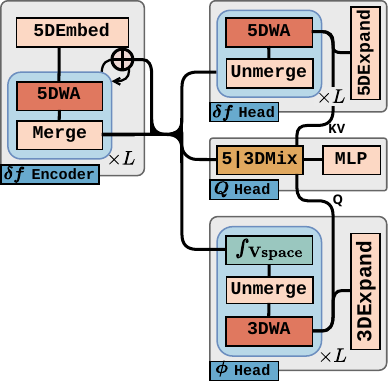}}    
        \caption{High-level \ourmethod{} architecture.}
        \label{fig:arch_gyroswin}
    \end{subfigure}%
    \begin{subfigure}{.60\textwidth}
        \centering
        \raisebox{15px}{\includegraphics[width=0.9\textwidth,
            trim=0 0 0 0,clip]{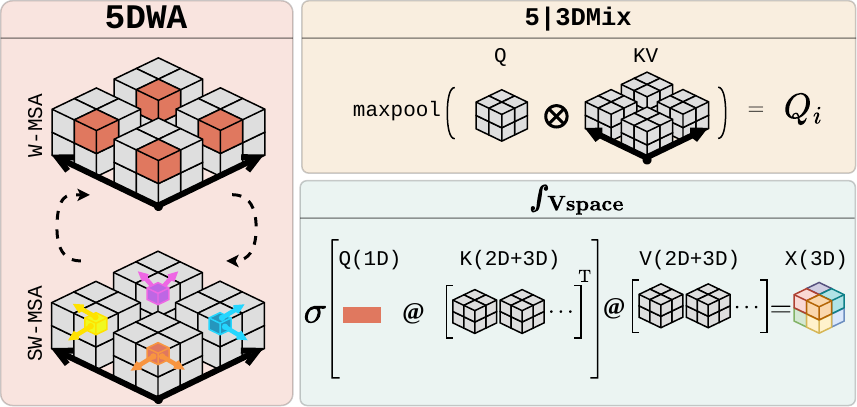}}   
        \caption{5D attention, latent mixing and integrator layers.}
        \label{fig:swin_attn}
    \end{subfigure}
    \caption{\textbf{Left:} \ourmethod{} receives the 5D distribution function as input and predicts the evolved 5D distribution function, as well as the respective 3D potential and heat flux at the next timestep. \textbf{Right:} Essential building blocks and integrator layers that enable multitask training. The latent 5D space is integrated over velocity space to obtain a latent 3D field for potential prediction via cross-attention.}
    \label{fig:gyroswin_arch}
\end{figure}

\textbfp{5D shifted window attention}
The main characteristic of Swin Transformer \citep{liu_swin_2021} is local attention in fixed window sizes.
This has a significant advantage over ViTs, reducing the quadratic complexity in token count to near-linear. 
More information on complexity for Swin and ViT can be found in \cref{app:swin_complexity}.
The core component of \ourmethod{} is 5DWA (see \cref{fig:swin_attn}).
In 5DWA, attention is performed across tokens within 5D windows of size $M := M_{v_{\parallel}} \times M_{v_\mu} \times M_s \times M_x \times M_y$.
Let $\BX \in \mathbb{R}^{b \times v_{\parallel} \times v_\mu \times s \times x\times y \times d}$ be a 5D field of batch size $b$ and hidden dimension $d$.
After partitioning $\BX$ into non-overlapping windows, we obtain $\BX_w \in \mathbb{R}^{\left(b \:\cdot\: N_{win}\right) \times \left(M_{v_{\parallel}}\cdot M_{v_{\mu}}\cdot M_{s}\cdot M_{x}\cdot M_{y}\right)\times d}$, with $N_{win} = v_{\parallel} / M_{v_{\parallel}} \cdot v_{\mu} / M_{v_{\mu}} \dots$ denoting the number of windows.
Self-attention is performed within each window in parallel across all windows. 
To enable interaction across neighboring windows we stack Window Multi-Head Attention (\texttt{W-MSA}) and Shifted Window MSA (\texttt{SW-MSA}) blocks.
This results in approximate global attention with increasing depth. 
We define \texttt{W-MSA} for any n-dimensional $x$ as
\begin{equation}
\label{attn}
    \texttt{W-MSA}(\BX) = \BW^{o}\left[\operatorname{SoftMax}\left(\frac{(\BX_w \BW^{Q}_h)(\BX_w \BW^{K}_h)^{\top}}{\sqrt{d}}\right) (\BX_w \BW^{V}_h) \right]_{h\in\text{heads}},
\end{equation}
where $\BW^{Q}_h, \BW^{K}_h, \BW^{V}_h$ are the head-wise query, key and value projection matrices for head $h$, and $\BW^{o}$ is the output matrix.
The shifted version \texttt{SW-MSA} is defined as spatially shifting the window partition stencil prior to \cref{attn}, and reversing the shift after
\begin{equation}
\label{swattn}
    \texttt{SW-MSA}(x) = \operatorname{roll}\left(\texttt{W-MSA}\left(\operatorname{roll}\left(\BX, -\frac{M}{2}\right)\right), +\frac{M}{2}\right),
\end{equation}
where $\operatorname{roll}\left(\BX, \pm\frac{M}{2}\right)$ is a matrix roll operation, which circularly shifts each dimension of $\Bx$ by $\pm\frac{M}{2}$ in the respective dimension.
For details on batched implementation of the cyclic shift we refer to \cite{liu_swin_2021} and \cite{kim2023swift}.

\textbfp{Up/Downsampling}
Following \citep{dosovitskiy_image_2021}, we employ \textit{patch embedding} to split the input into non-overlapping patches via 5D local convolution and embed them into tokens.
The inputs are then spatially downsampled by the patch size and channels are projected to dimension $C$. 
The downsampling path of the UNet interleaves 5DWA blocks with \textit{patch merging} layers to compress the 5D field at each stage. 
We extend these patch merging layers from \citet{liu_swin_2021} to 5D: they concatenate features of $2\times 2 \times 2 \times 2 \times 2$ neighboring patches and apply an MLP, reducing resolution by $2^{5}$ times. 
The output dimension is set to $2C$, which doubles the channels after each merge.
We store intermediate feature maps for skip connections to produce hierarchical representations at different resolutions, resulting in growing receptive field and lower computational cost at lower stages.
For upsampling we employ \textit{patch expansion} layers, reversing the patch embedding (and merging) via a linear projection and rearrangement to the original field size.
\cref{fig:arch_gyroswin} sketches the architecture of \ourmethod{}, incorporating patch embedding/merging/expansion blocks along with 5DWA.

\textbfp{Multitask training}
To ensure that \ourmethod{} adheres to downstream integrated quantities, such as electrostatic potential fields $\bm{\phi}$ and scalar flux $Q$, we add prediction tasks for each of them in addition to predicting $f$.
The $\bm{\phi}$-head is constructed of $L$ blocks, each of which employs an integral layer to reduce the 5D latent to a 3D latent followed by 3DWA layers and expansion layers until the original spatial resolution is recovered.
For the $Q$-head, we perform max pooling over the integrated 3D space followed by an MLP head.
Importantly, we also employ latent $\text{5D}\xleftrightarrow{}\text{3D}$ mixing layers at each block to facilitate latent communication among the different heads.
For multitask training of \ourmethod{}, we introduce loss terms for each head with weighting factors:
\begin{equation}\label{eq:losses}
    \mathcal{L} = w_{f}\mathcal{L}_f(\hat{f}, f) + w_\phi\mathcal{L}_\phi(\hat{\phi}, \phi) + w_{\bar{Q}}\mathcal{L}_{\bar{Q}}(\hat{Q}, \bar{Q}) . 
\end{equation}

\textbfp{5D$\xleftrightarrow{}$3D mixing and integrator modules}
Multitask training on 3D potential fields requires reducing the 5D latent space to a latent 3D space.
To this end, we incorporate latent integrator modules (see $\boldsymbol{\smallint}_{\textbf{Vspace}}$ block in \cref{fig:swin_attn}).
They perform cross-attention between a learnable 1D query $\BQ \in \mathbb{R}^{1 \times d}$, which is broadcasted over the velocity space of $\BK,\BV \in \mathbb{R}^{(b \cdot s \cdot x\cdot y) \times v_{\parallel} \times \mu \times d}$.
This procedure is reminiscent of perceiver-style pooling \citep{jaegle_perceiver_2021} with a single fixed query across velocity dimensions.
Moreover, we allow interaction between 3D and 5D latents via cross-attention in each block $L$, where $\BQ \in \mathbb{R}^{b \times s \times x\times y \times d}$ and $\BK, \BV \in \mathbb{R}^{b \times (v_{\parallel} \cdot v_\mu) \times s \times x\times y \times d}$.
\cref{fig:swin_attn} shows an example of latent $\text{3D}\rightarrow\text{5D}$ interaction.
by swapping $\BQ$ with $\BK$ and $\BV$, we also obtain $\text{5D}\rightarrow\text{3D}$ cross-attention.
We stack both cross-attention layers to obtain our \textit{5|3DMix} bloks.

\section{Experiments}
\label{sec:experiments}

In this section we elaborate on our experimental setup, ranging from data generation to baselines and our evaluation setup.
For implementation details see \cref{app:impl_details}.

\textbfp{Data Generation}
We run nonlinear simulations using the numerical code GKW \citep{PEETERS_GKW_2009}, varying noise amplitude of initial conditions and four operating parameters, namely the safety factor $q$, the magnetic shear $\hat{s}$, the ion temperature gradient $R/L_t$ and the density gradient $R/L_n$.
To reduce the computational burden of data generation, we consider the adiabatic electron approximation at a resolution of $(32 \times  8 \times 16 \times 85 \times 32)$, i.e. we only consider turbulence originating from ion temperature gradients.
We use latin hypercube sampling to uniformly populate the parameter space \citep{McKay_lhs_1979} and chose parameter regions to ensure that the resulting simulations are highly turbulent.
The distribution of operating parameters and corresponding heat flux can be observed in \cref{fig:param_dist}.
We run each simulation for a total of 31,920 steps, which are averaged every 40 steps and subsampled every third step, resulting in a total of 266 snapshots.
Each snapshot comes with two channels, representing the real and imaginary parts of the ballooning transform, commonly used for plasma coordinates.
We neglect the first 80 snapshots of each simulations as those correspond to the linear phase where turbulence is not fully established yet.
The entire dataset comprises 255 simulations, based on which we assemble two training subsets, one comprising 48 simulations and another one comprising 241 simulations.
We use the small subset for comparison to baselines and the latter for scaling experiments.
In total, this results in 44{,}400 training samples.
We provide visualizations of sample snapshots of the 5D fields in \cref{app:data_generation_visualisation}.

\textbfp{Baselines}
We implement three types of baselines:
\begin{enumerate*}[label=(\textbf{\roman*})]
\item As a state-of-the-art ROM, we implement the QuaLiKiz saturation rule \citep{Bourdelle2015} and fit it to nonlinear fluxes of our training set, similarly to \citet{Kumar_qlgkw_2021}.
For details, see \cref{app:quasilinear}.
\item Tabular regressors, such as Gaussian Process Regression (GPR) for nonlinear fluxes
as proposed by \citep{Hornsby2024}, and a MLP trained in the same manner, akin to \citet{plassche_qlknn_2020}.
\item Neural surrogates, trained to predict the 5D density function:
Fourier Neural Operator \citep[FNO]{li2021fourierneuraloperatorparametric}, PointNet \citep{qi2016pointnet}, Transolver \citep{wu2024Transolver}, and vanilla ViT \citep{dosovitskiy_image_2021}.
\end{enumerate*}
Additional information on baselines can be found in \cref{app:impl_details}.

\textbfp{Evaluation}
The promise of \ourmethod{} is that it can replace QL approaches as it is trained on the full 5D distribution function, but maintains efficiency and scalability.
To properly evaluate whether our \ourmethod{} yields improvements over QL models, we compile a set of nonlinear simulations in a high ion temperature gradient regime, ensuring a strong turbulence regime.
We set aside 14 of the 255 simulations that we generate in total to evaluate for in-distribution (\textbf{ID}) and out-of-distribution (\textbf{OOD}) generalization. 
For the \textbf{ID} set, we identify a region in the 4D parameter set that lives within the convex hull of the training set, but is unseen to the model.
Conversely, we ensure that the \textbf{OOD} set is not governed by the convex hull of the training simulations.
In total, we compile select six simulations for the \textbf{ID} set and five simulations for the \textbf{OOD} set.
The remaining three simulations are used as a validation set during training.
All of them are excluded from the training set.

\section{Results}
\label{sec:results}
As shown in \cref{tab:surrogate_capabilities} different methods are restricted to certain evaluations, i.e. tabular regressors can only be evaluated on nonlinear flux prediction (scalars), while QL models can additionally be used to evaluate for diagnostics.
Neural surrogates that predict the full 5D field, such as \ourmethod{}, are the only class that can also be evaluated for zonal flows.
In line with this observation, we provide results for each of the three categories based on their capabilities.

\begin{table}[ht]
  \caption{Comparison of different surrogate approaches by capabilities.}
  \label{tab:surrogate_capabilities}
  \centering
  \setlength{\tabcolsep}{8pt}
  \renewcommand{\arraystretch}{1.2}
  \resizebox{0.9\columnwidth}{!}{
  \begin{tabular}{lcccc}
    \toprule
    \textbf{Method} & \textbf{Average Flux} & \textbf{Diagnostics} & \textbf{Zonal Flows} & \textbf{Turbulence} \\
    \midrule
    Tabular Regressors, e.g., GPR, MLP         & \cellcolor{green!20}\textbf{1D$\rightarrow$0D} & \xmark & \xmark & \xmark \\
    SOTA Reduced Numerical modelling, e.g., QL    & \cellcolor{green!20}\textbf{3D$\rightarrow$0D} & \cellcolor{green!20} \textbf{3D$\rightarrow$1D} & \xmark  & \xmark \\
    Neural Surrogates, e.g. \ourmethod{} (Ours) & \cellcolor{green!20}\textbf{5D$\rightarrow$0D} & \cellcolor{green!20} \textbf{5D$\rightarrow$1D} & \cellcolor{green!20}\textbf{5D$\rightarrow$1D} & \cellcolor{green!20}\textbf{5D$\rightarrow$5D} \\
    \bottomrule
  \end{tabular}}
\end{table}

\textbfp{5D$\rightarrow$5D Turbulence modelling}
To evaluate for 5D turbulence modelling capabilities, we perform an autoregressive rollouts with the neural surrogates and measure correlation time.
Following \citet{alkin_upt_2024}, we define correlation time as the number of snapshots that can be predicted while maintaining a certain pearson correlation $\tau$.
This metric, demonstrates what methods are capable of performing stable autoregressive rollouts without drifting too far from the ground truth.
We report correlation times for the neural surrogates in \cref{tab:main_results}.
We observe a clear trend that \ourmethod{} is by far the most stable autoregressive method compared to other neural surrogates.
Remarkably, when scaling \ourmethod{} in terms of data and model size, we attain stable rollouts for over 100 timesteps, even for \textbf{OOD} simulations.

\textbfp{5D$\rightarrow$0D average flux}
The task for this evaluation is to predict the average heat flux $\bar{Q}$ over the last 80 timesteps of a simulation from the full 5D field after an autoregressive rollout.
We present results for both ID and OOD evaluation sets in \cref{tab:main_results}.
On the reduced training set \ourmethod{} yields the best performance compared to the QL model and alternative 5D neural surrogates.
Interestingly, most neural surrogates converge to similar error for $\bar{Q}$ ($\sim 120$).
This error is obtained if magnitudes of predictions decay along the rollout resulting in $\bar{Q} \sim 0$ and therefore error of $\sim 120$.
This indicates that all competitors fail to maintain turbulent transport along the rollout as flux predictions decay to zero, leading to similar errors in $\bar{Q}$.
Furthermore, when scaling \ourmethod{} to more data and larger model sizes, we observe a drastic improvement in nonlinear flux prediction, achieving significantly lower error than currently existing surrogates.
In \cref{app:ablation_studies} we also show that the composition of all components of \ourmethod{} results in the best performance.

\begin{table}[]
    \caption{Evaluation for 5D turbulence modelling and nonlinear heat flux prediction. We evaluate all methods across six in-distribution (\textbf{ID}) and five out-of-distribution (\textbf{OOD}) simulations. %
    For $\bar{Q}$ we report RMSE of time-averaged predictions after an autoregressive rollout. For $f$ we report correlation time for autoregressive rollouts with threshold $\tau=0.1$. Higher correlation time is better. 
    }
    \label{tab:main_results}
    \centering
      \setlength{\tabcolsep}{10pt}
      \renewcommand{\arraystretch}{1.1}
      \resizebox{\textwidth}{!}{
      \begin{tabular}{l|c|cc|cc}
        \toprule
        \multirow{2}{*}{\textbf{Method}} & \multirow{2}{*}{\textbf{Input}} & \multicolumn{2}{c|}{$f$} & \multicolumn{2}{c}{$\bar{Q}$}  \\ %
        & & \textbf{ID} ($\uparrow$) & \textbf{OOD} ($\uparrow$) & \textbf{ID} ($\downarrow$) & \textbf{OOD} ($\downarrow$) \\
        \midrule
        \multicolumn{6}{c}{\textit{SOTA Reduced Numerical modelling}} \\
        \midrule
        QL \citep{bourdelle_qualikiz_2007} & 
        3D & n/a & n/a & 89.48 $\pm$ 11.77 & 95.18 $\pm$ 21.58 \\ %
        \midrule
        \multicolumn{6}{c}{\textit{Classical Regression Techniques}} \\
        \midrule
        GPR \citep{Hornsby2024}  & 0D & n/a & n/a & 43.82 $\pm$ 10.84 & 59.28 $\pm$ 17.55 \\
        MLP & 0D & n/a & n/a & 50.90 $\pm$ 10.87 & 62.88 $\pm$ 18.58 \\
        \midrule
        \multicolumn{6}{c}{\textit{Neural Surrogate Models (48 simulations)}} \\
        \midrule
        FNO \citep{li2021fourierneuraloperatorparametric} & 3D & 9.33 $\pm$ 0.56 & 9.20 $\pm$ 0.58 & 119.88 $\pm$ 13.15 & 124.96 $\pm$ 23.27 \\
        PointNet \citep{qi2016pointnet} & 5D & 8.5 $\pm$ 1.26 & 9.0 $\pm$ 0.89 & 119.90 $\pm$ 13.16 & 125.01 $\pm$ 23.28 \\
        Transolver \citep{wu2024Transolver} & 5D & 9.66 $\pm$ 0.94 & 10.2 $\pm$ 0.75 & 119.92 $\pm$ 13.15 & 125.02 $\pm$ 23.28 \\
        ViT \citep{dosovitskiy_image_2021} & 5D & 16.83 $\pm$ 1.49 & 19.20 $\pm$ 1.36 & 119.63 $\pm$ 13.13 & 125.13 $\pm$ 23.29 \\ 
        \ourmethod{} (Ours) & 5D & 17.67 $\pm$ 2.16 & 19.40 $\pm$ 1.46 & 47.35 $\pm$ 6.20	& 59.15 $\pm$ 15.62 \\
        \midrule
        \multicolumn{6}{c}{\textit{Scaling \ourmethod{} to 241 simulations}} \\
        \midrule
        \ourmethodsmall{} (Ours) &  5D & 49.00 $\pm$ 4.62 & 74.80 $\pm$ 23.71 & 16.37 $\pm$ 3.68 & 16.73 $\pm$ 5.58 \\
        \ourmethodmedium{} (Ours) & 5D & 95.50 $\pm$ 20.34 & 74.60 $\pm$ 11.10 & 20.41 $\pm$ 3.86 & 24.96 $\pm$ 5.63 \\
        \ourmethodlarge{} (Ours) & %
        5D & \textbf{110.33} $\pm$ 19.74 & \textbf{111.80} $\pm$ 23.86 & 18.35 $\pm$ 1.56 & 26.43 $\pm$ 9.49 \\ %
        \bottomrule
      \end{tabular}
      }
\end{table}

\textbfp{Scalability} 
In integrated plasma simulations, turbulent heat fluxes must be repeatedly computed across thousands of runs, radial locations, and time intervals, underscoring the need for scalable surrogate architectures.
To assess scalability of neural surrogates, we report inference speed, memory consumption, and number of parameters on a single NVIDIA H100 80GB HBM3 in \cref{tab:main_diagnostics}.
We exclude the 3D FNO as it is based on collapsing velocity dimensions into channels (see \cref{app:impl_details}) which does not scale to higher resolutions.
Factorized variants of FNO and CNN are also slow and memory heavy.
Field-based surrogates (PointNet, Transolver) require subsampling during training and chunked inference, limiting scalability.
ViTs suffer from quadratic complexity and similar memory consumption as \ourmethod{} despite using half of its parameters.
Scaling ViT to the same parameter count as \ourmethodsmall{} results in $\sim18\text{ms}$ inference speed which is $52.5\%$ slower than \ourmethodsmall{}. 
\ourmethodsmall{} is roughly three orders of magnitude faster than the numerical code GKW ($4200$ vs. $756$ GFLOPs).

\begin{figure}
    \centering
    \includegraphics[width=\linewidth]{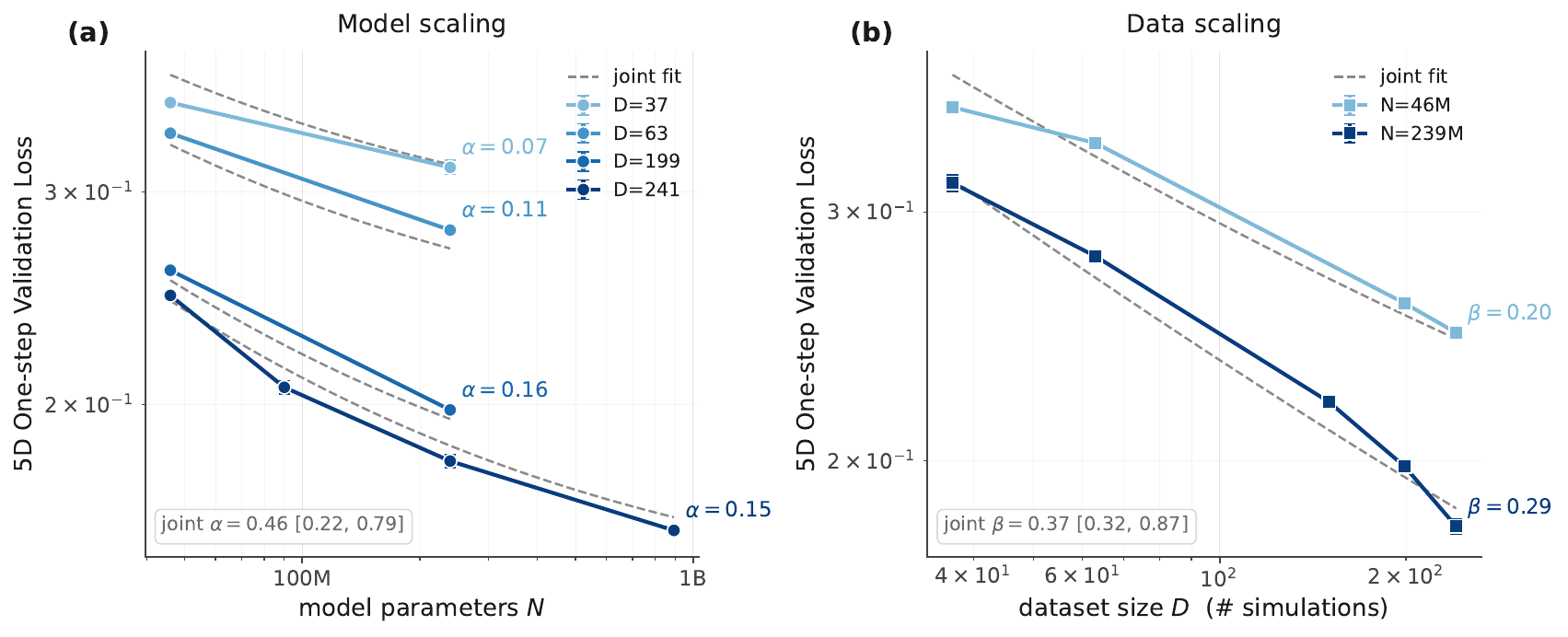}
    \caption{Model scaling \textbf{(a)} and data scaling laws \textbf{(b)} for \ourmethod{}. We train four model sizes ranging from 46M to 893M parameters and varying numbers of simulations from 37 to the full dataset of 241. Scaling \ourmethod{} to 893M parameters enters the tail of the scaling law, while data scaling does not enter saturation.}
    \label{fig:gyroswin_scaling}
\end{figure}

Furthermore we report scaling laws to determine the compute-optimal size of \ourmethod{} for the data we generated \citep{hoffmann2022training}.
Across the $11$ $(N,D)$ configurations in \cref{fig:gyroswin_scaling}, the one-step validation loss follows a joint power law $L(N,D)=E+A\,N^{-\alpha}+B\,D^{-\beta}$ with $\alpha=0.46$ (95\% CI $[0.22,0.79]$) and $\beta=0.37$ ($[0.32,0.87]$).
The irreducible term $E$ is not identifiable from our data, i.e.\ we observe no saturation. 
Increasing the model from 239M to 893M parameters at full data still reduces the loss by $12\%$, and the seed-to-seed spread ($\leq 0.004$) is far below every axis effect.
The two axes are strongly coupled as the local model-size exponent grows from $0.07$ at $37$ simulations to $0.15$--$0.16$ at $\geq 199$ simulations, and the local data exponent grows from $0.20$ for the 46M model to $0.29$ for the 239M model.
Small model variants cannot exploit additional data and small datasets cannot exploit additional capacity, so model size and dataset size must be scaled together.

\textbfp{5D$\rightarrow$1D flux spectrum}
A major advantage of quasilinear approaches over tabular regressors is physical verifiability.
As they are based on 3D linear simulations, flux contributions for each mode in $k_y$ ($Q(k_y)$) from \cref{eq:q_spectrum} can be inspected. 
This provides insights as to whether modes are captured that contribute most to heat transport.
Similarly, we assess whether the general structure of $Q(k_y)$ is captured in predicted 5D snapshots of neural surrogates.
We report pearson correlation to the ground-truth $Q(k_y)$ for all 5D neural surrogates in \cref{tab:main_diagnostics}.
We observe that all neural surrogates exhibit higher correlation than the QL approach highlighting their potential.
Furthermore, we demonstrate that \ourmethod{} exhibits the best correlation on the \textbf{OOD} test cases.
However, it performs slightly worse on the \textbf{ID} cases than competitors.
When scaling \ourmethod{} to the large training set, we find a boost in performance, resulting in almost perfect reproduction of the shape of $Q(k_y)$.
This means that \ourmethodlarge{} captures the energy transport per-mode almost perfectly.
In \cref{fig:all_fluxspecs} we visualize $Q(k_y)$ for all \textbf{ID} as well as \textbf{OOD} cases.

\begin{figure}
    \centering
    \includegraphics[width=\linewidth]{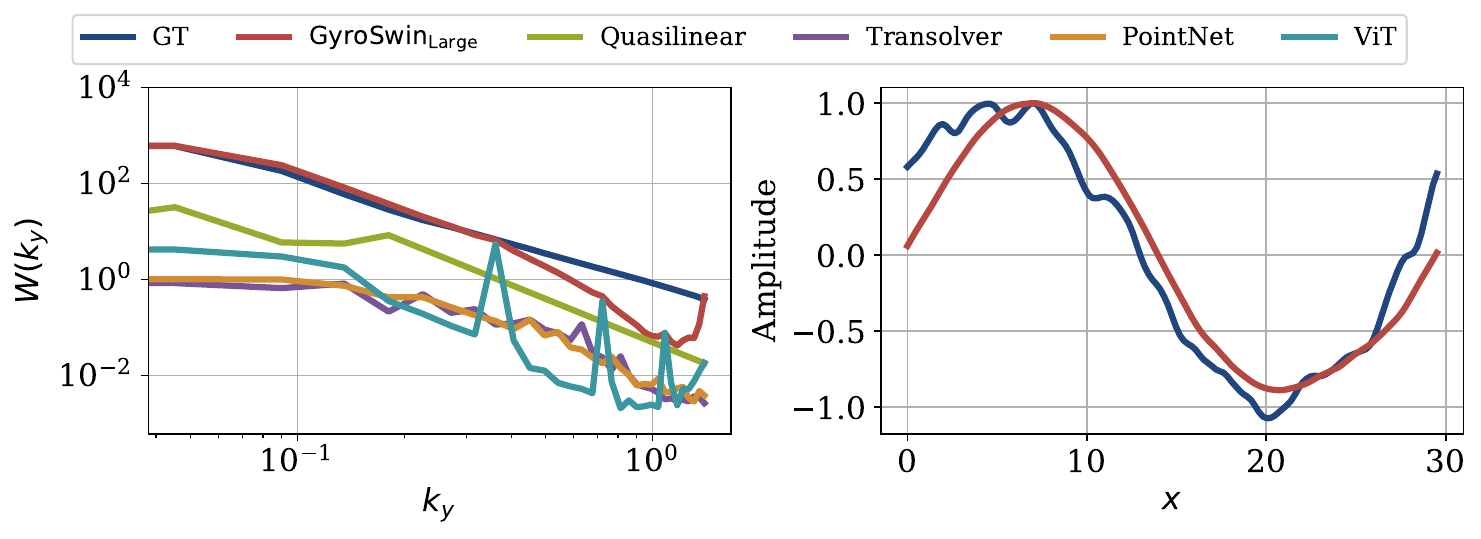}
    \caption{\textbf{Left:} $W(k_y)$ averaged over time and \textbf{OOD} simulations for different 5D neural surrogates. Competitors tend to underestimate while \ourmethod{} matches the spectrum well with a slight discrepancy on higher frequencies. \textbf{Right:} Time-averaged zonal flow profile for a slice along $s$ across radial coordinates $x$ for a selected \textbf{OOD} simulation. \ourmethod{} captures the zonal flow profile.}
    \label{fig:kyspec_zf}
\end{figure}

As mentioned in \cref{sec:backgrnd} another useful diagnostic is the turbulence intensity spectrum $W(k_y)$ which is not necessarily aligned with $Q(k_y)$, i.e. the mode transporting the most energy might not be the same where turbulence is most intense.
Therefore we visually showcase $W(k_y)$ in \cref{fig:kyspec_zf} (left) averaged across time and all \textbf{OOD} cases.
In addition we provide visualizations for each test case of \textbf{ID} and \textbf{OOD} separately in \cref{fig:all_kyspecs}.
We observe that while all neural surrogates decently reproduce the shape of the spectrum, they heavily underestimate the magnitude.
Furthermore, \ourmethodlarge{} matches shape and magnitude almost perfectly, up until a few higher frequency components.
We attribute this finding to the spectral bias of neural networks towards lower frequency (higher energy) modes \citep{rahaman_spectral_2019}.

\begin{table}[]
    \caption{Evaluation for scalability, diagnostics, and nonlinear physics. We evaluate all methods across six in-distribution (\textbf{ID}), and five out-of-distribution (\textbf{OOD}) simulations. 
    For diagnostics and nonlinear physics we report time-averaged pearson correlation for autoregressive rollout.
    \ourmethod{} is the only method that can be scaled to $\sim$1B parameters while maintaining reasonable inference speed and improving correlation to diagnostics.}
    \label{tab:main_diagnostics}
    \centering
      \setlength{\tabcolsep}{10pt}
      \renewcommand{\arraystretch}{1.1}
      \resizebox{\textwidth}{!}{
      \begin{tabular}{l|ccc|cc}
        \toprule
        \multirow{2}{*}{\textbf{Method}} & \multirow{2}{*}{\textbf{Fwd[ms]}} & \multirow{2}{*}{\textbf{Mem[GB]}} & \multirow{2}{*}{\textbf{Params[M]}}  & %
        \multicolumn{2}{c}{$Q(k_y)$} \\ %
        & & & & \textbf{ID} ($\uparrow$) & \textbf{OOD} ($\uparrow$) \\
        \midrule
        \multicolumn{6}{c}{\textit{SOTA Reduced Numerical modelling}} \\
        \midrule
        QL \citep{bourdelle_qualikiz_2007} & N/A & N/A & N/A &%
        0.51 $\pm$ 0.4 & 0.58 $\pm$ 0.33 \\
        \midrule
        \multicolumn{6}{c}{\textit{Neural Surrogate Models (48 simulations)}} \\
        \midrule
        F-CNN & 569.6 & 11.1 & 2.0 & N/A & N/A  \\
        F-FNO & 963.3 & 36.9 & 1.3 & N/A & N/A  \\
        PointNet \citep{qi2016pointnet} & 19.7 & 39.7 & 43.9 & %
        0.49 $\pm$ 0.02 & 0.54 $\pm$ 0.09 \\ %
        Transolver \citep{wu2024Transolver} & 3103 & 35.8 & 40.6 & %
        0.50 $\pm$ 0.02 & 0.53 $\pm$ 0.09 \\ %
        ViT \citep{dosovitskiy_image_2021} & 6.3 & 2.4 & 46.1 & %
        0.52 $\pm$ 0.05 & 0.63 $\pm$ 0.11 \\ %
        \ourmethod{} (Ours) & 11.8 & 2.8 & 90.2 & %
        0.59 $\pm$ 0.06 & 0.69 $\pm$ 0.06 \\ %
        \midrule
        \multicolumn{6}{c}{\textit{Scaling \ourmethod{} to 241 simulations}} \\
        \midrule
        \ourmethodsmall{} (Ours) & 11.8 & 2.8 & 90.2 & 
        0.82 $\pm$ 0.10 & 0.79 $\pm$ 0.09 \\ %
        \ourmethodmedium{} (Ours) & 12.1 & 5.3 & 250.9 &
         0.72 $\pm$ 0.08 & 0.76 $\pm$ 0.12 \\ %
        \ourmethodlarge{} (Ours) & 
        15.4 & 9.6 & 998.3 & %
        \textbf{0.87} $\pm$ 0.11 & \textbf{0.91} $\pm$ 0.08 \\ %
        \bottomrule
      \end{tabular}
      }
\end{table}

\textbfp{New capability: 5D zonal flow modelling}
Zonal flows have a significant impact on the turbulence dynamics.
To evaluate whether \ourmethod{} is capable of capturing zonal flows, we visualize the zonal flow profile of a test case of the \textbf{OOD} set.
We again time-average the profiles and compare to the time-averaged predicted zonal flow profile of \ourmethodlarge{}.
We find that \ourmethodlarge{} accurately captures the zonal flow profile.
This capability has so far been unreachable by any other surrogate modelling technique.
We provide additional visualizations for each test case of the \textbf{ID} and \textbf{OOD} sets in \cref{fig:all_zfprofiles}.

\section{Conclusions and Limitations}
\label{sec:conclusion}

We present \ourmethod{}, a scalable neural surrogate model for nonlinear gyrokinetic equations modelling turbulent transport in Plasmas.
Unlike existing surrogate models, \ourmethod{} operates directly in a 5D space and evolves the 5D distribution function of gyrokinetics.
It is based on latent cross-attention and integration modules for 3D$\xleftrightarrow{}$5D latent interaction and trained in a multitask fashion to predict the distribution function, electrostatic potentials and heat fluxes.
Furthermore, we perform channelwise mode separation to incorporate an inductive bias towards essential nonlinear phenomena (zonal flows) that are essential for modelling turbulence.
We show that \ourmethod{} outperforms reduced numerical appraoches and scales favorably compared to other neural surrogates.
Furthermore, \ourmethod{} accurately captures nonlinear phenomena self-consistently, a capability not present in prior surrogate or reduced numerical models.
As a result, \ourmethod{} offers a fruitful alternative to efficient approximation of turbulent transport.

Currently, the main limitation of \ourmethod{} is that it does not take into account the chaotic and therefore distributional nature of turbulence. 
Although we find that \ourmethod{} produces stable autoregressive rollouts over 100 timesteps, it suffers from error accumulation.
Generative modelling offers a remedy to this problem by directly predicting snapshots in the saturated phase. 
We aim to incorporate such distributional approaches in future work.
Furthermore, we neglect the linear phase of the simulation.
The reason for this is that our focus primarily lies in modelling turbulence.
In the future we aim to extend \ourmethod{} to the linear phase as well.
Finally, we only consider the adiabatic electron approximation, as it allows generation of a relatively large training set at rather low cost.
Still, each simulation produces considerable data volumes, making full coverage of the 4D parameter space difficult.
Beyond that, we envision a high-fidelity surrogate model by transfer learning from low-fidelity approximate simulations.

\begin{ack}
We extend special thanks to our colleagues Wei Lin for the discussion on computer vision and multi-task learning, and Anna Zimmel and Gerald Gutenbrunner for conversations on the spectral behavior of neural networks.

This work has been funded by the Fusion Futures Programme. As announced by the UK Government in October 2023, Fusion Futures aims to provide holistic support for the development of the fusion sector.
The ELLIS Unit Linz, the LIT AI Lab, the Institute for Machine Learning, are supported by the Federal State Upper Austria. We thank the projects FWF AIRI FG 9-N (10.55776/FG9), AI4GreenHeatingGrids (FFG- 899943), Stars4Waters (HORIZON-CL6-2021-CLIMATE-01-01), FWF Bilateral Artificial Intelligence (10.55776/COE12). 
We acknowledge EuroHPC Joint Undertaking for awarding us access to Leonardo at CINECA, Italy, and Deucalion at MACC, Portugal.
Parts of the computational results have been achieved using the Austrian Scientific Computing (ASC) infrastructure.
\end{ack}

\bibliographystyle{neurips_2025}
\bibliography{references}

\newpage
\appendix

\section{Derivation of the Gyrokinetic equation}
\label{app:gyrokinetics}

We begin with the Vlasov equation for the distribution function $f(\Br, \Bv, t)$:
\begin{equation}
\frac{\partial f}{\partial t} + \mathbf{v} \cdot \nabla f + \frac{q}{m} \left( \BE + \Bv \times \BB \right) \cdot \nabla_v f = 0
\end{equation}

The Vlasov equation describes the conservation of particles in phase space in the absence of collisions.
Here, $\Br=(x, y, s)$ and $\Bv=(v_x, v_y, v_s)$ correspond to coordinates in the spatial and the velocity domain, respectively.
Hence the Vlasov equation is a 7D (including time) PDE  representing the  density of particles in phase space at position $\Br$, velocity $\Bv$, and time.
The term $\nabla_{\Bv} f$ describes the response of the distribution function to accelerations of particles and $\frac{q}{m} \left( \mathbf{E} + \mathbf{v} \times \mathbf{B} \right)$ denotes the Lorentz force, which depends on particle charge $q$ and mass $m$, as well as electric field $\BE$ and magnetic field $\BB$.
Finally, the advection (or convection) term $\Bv \nabla f$ describes transport of the distribution functon through space due to velocities.

To derive the \textit{gyrokinetic equation}, we transform from particle coordinates to guiding center coordinates $(\mathbf{R}, v_\parallel, \mu, \theta)$, where $\mu = \frac{m v_\perp^2}{2B}$ is the magnetic moment, $\theta$ the gyrophase, which describes the position of a particle around its guiding center as it gyrates along a field line, and $\BR$ is the coordinate of the guiding center.

Assuming the time scale  $L$ at which the background field changes is much longer than the gyroperiod with a small Larmor radius $\rho \ll L$, we can \textit{gyroaverage} to remove the dependency on the gyrophase $\theta$, yielding:
\begin{equation}
\frac{\partial f}{\partial t} + \dot{\BR} \cdot \nabla f + \dot{v}_\parallel \frac{\partial f}{\partial v_\parallel} = 0
\end{equation}

\subsection*{Linear Terms}

The unperturbed (background) motion of the guiding center is governed by:
\begin{align}
\dot{\BR} &= v_\parallel \Bb + \Bv_D \\
\dot{v}_\parallel &= -\frac{\mu}{m} \Bb \cdot \nabla \BB
\end{align}

Here,  $\mathbf{b} = \BB/B$ is the unit vector along the magnetic field, and $\Bv_D$ represents magnetic drifts. Substituting into the kinetic equation yields
\begin{equation}
\frac{\partial f}{\partial t} + (v_\parallel \Bb + \Bv_D) \cdot \nabla f 
- \frac{\mu}{m} \Bb \cdot \nabla \BB \frac{\partial f}{\partial v_\parallel} = 0
\end{equation}

We can express the magnetic gradient term using:
\begin{equation}
    \Bb \cdot \nabla \BB = \frac{\BB \cdot \nabla \BB}{B}    
\end{equation}
so that:
\begin{equation}
\frac{\mu}{m} \Bb \cdot \nabla \BB = \frac{\mu B}{m} \frac{\BB \cdot \nabla B}{\BB^2}
\end{equation}

\subsection*{Nonlinear Term}

Fluctuating electromagnetic potentials $\delta \phi, \delta \BA$ induce E$\times$B and magnetic flutter drifts. 
We define the gyroaveraged generalized potential as
\begin{equation}
    \chi = \langle \phi - \frac{v_\parallel}{c} \BA_\parallel \rangle,    
\end{equation}
where $\BA_\parallel$ is the aprallel component of the vector potential, $\langle\cdot\rangle$ denotes the gyroaverage, and $c$ is the speed of light, which is added to ensure correct units. $\phi$ is the electrostatic potential, the computation of which involves an integral of $f$ over the velocity space (see eq. 1.41 in the GKW manual \footnote{\url{https://bitbucket.org/gkw/gkw/src/develop/doc/manual/}} for a complete description). 

This gives rise to the drift
\begin{equation}
    \mathbf{v}_\chi = \frac{c}{B} \mathbf{b} \times \nabla \chi, 
\end{equation}
and yields the nonlinear advection term $\mathbf{v}_\chi \cdot \nabla f$ .

\subsection*{Final Equation}

We arrive at the gyrokinetic equation in split form:
\begin{equation}
\boxed{
\underbrace{\colorbox{blue!10}{$\displaystyle
\frac{\partial f}{\partial t} + (v_\parallel \mathbf{b} + \mathbf{v}_D) \cdot \nabla f   
-\frac{\mu B}{m} \frac{\mathbf{B} \cdot \nabla B}{B^2} \frac{\partial f}{\partial v_\parallel}$}}_\text{Linear}
\; + \;
\underbrace{\colorbox{red!30}{$\displaystyle
\mathbf{v}_\chi \cdot \nabla f$}}_\text{Nonlinear}
\; = \;
S
}
\end{equation}

Here, \( S \) represents external sources, collisions, or other drive terms.
To enhance the tractability of \cref{gyrovlas}, the distribution function $f$ is usually split into equilibrium and perturbation terms
\begin{equation}
    f = f_0 + \delta f = \underbrace{\colorbox{blue!10}{$\displaystyle f_0 - \frac{Z\phi}{T}f_0$}}_\text{Adiabatic}\; + \; \underbrace{\colorbox{red!30}{$\displaystyle \frac{\partial h}{\partial t}$}}_\text{Kinetic} \ ,
\end{equation}
where $f_0$ is a background or equilibrium distribution function, $T$ the particle temperature, $Z$ the particle charge, $\phi$ the electrostatic potential, and $\delta f$ the total perturbation to the distribution function, which comprises of \textit{adiabatic} and \textit{kinetic} response. 
The adiabatic term describes rapid and passive responses to the electrostatic potential that do not contribute to turbulent transport, while the kinetic term governs irreversible dynamics that facilitate turbulence. Numerical codes, such as GKW \citep{PEETERS_GKW_2009}, rely on solving for $\delta f$ instead of $f$.
A common simplification is to assume that electrons are adiabatic, which allows us to neglect the kinetic term in the respective $\delta f$.
Hence, the respective $f$ for electrons ($f_e$) does not need to be modelled, effectively halving the computational cost.

\section{Quasilinear models}
\label{app:quasilinear}

We used the QuaLiKiZ saturation rule \citep{bourdelle_qualikiz_2007} applied to lienar GKW simulations, following \citet{Kumar_qlgkw_2021}.
The QuaLiKiz saturation rule estimates turbulent transport based on linear gyrokinetic stability analysis, using quasilinear theory and empirical saturation rules. The quasilinear ion heat flux $Q$ is modelled as:

\begin{equation} \label{qlk_weighted_sum}
    Q = \sum A_k W_k
\end{equation}

where $A_k$ is the \textit{linear weight spectrum} that quantifies the phase relationship between electrostatic potential fluctuations and that is retrieved by solving the linear gyrokinetic equation, and $W_k$ is the \textit{turbulence intensity spectrum} which in QuaLiKiz is parametrised by a shape function $S_k$ and a normalisation factor $C$ calibrated on nonlinear gyrokinetic simulations.

To fit the normalisation factor $C$, we assume access to a vector of nonlinear fluxes $\Bq \in \mathbb{R}^j$ and set the right-hand side of \cref{qlk_weighted_sum} to $\Bx \in \mathbb{R}^j$, then the optimal solution for $C$ can be obtained via a least-squares fit
\begin{equation}
    C = \frac{\sum_j x_j q_j}{\sum_j x_i^2}.
\end{equation}

In our setup, we used flux estimates from the small training set (48 simulations) to compute the fit that resulted in a $C=7.93$.
This value differs slightly from the one reported in \citet{Kumar_qlgkw_2021}.
We attribute the difference to the different sampling strategy of simulations.
The QuaLiKiz saturation rule retains key physics from gyrokinetics while allowing efficient prediction of turbulent transport in integrated modelling frameworks.
Furthermore, we found it beneficial to center the nonlinear flux vector $\Bq$ which alters the normalisation constant $C$, but leads to reduced error on flux predictions on a separate validation set.

\section{Data Generation and Visualisation}
\label{app:data_generation_visualisation}

\textbfp{Data Generation} 

In this work, we mainly consider turbulence driven by ion temperature gradients, also typically referred to as the cyclone base case \citep{dimits_cbc_2000}.
We leverage the numerical code GKW \citep{PEETERS_GKW_2009} for generating nonlinear and linear gyrokinetic simulations, by varying four parameters: $R/L_t$, $R/L_n$, $\hat{s}$, and $q$, which significantly affect emerging turbulence in the Plasma.
$R/L_t$ is the ion temperature gradient, which is the main driver of turbulence.
$R/L_n$ is the density gradient, which has a more stabilizing effect but can sometimes also lead to enhanced turbulence.
The parameter $q$ denotes the so-called safety factor, which is the inverse of the rotational transform and describes how often a particle takes a poloidal turn before taking a toroidal turn.
$\hat{s}$ denotes magnetic shearing. 
It has a stabilizing effect as more magnetic shearing results in improved isolation of the Plasma.

\begin{figure}[t]
    \centering
    \includegraphics[width=\textwidth]{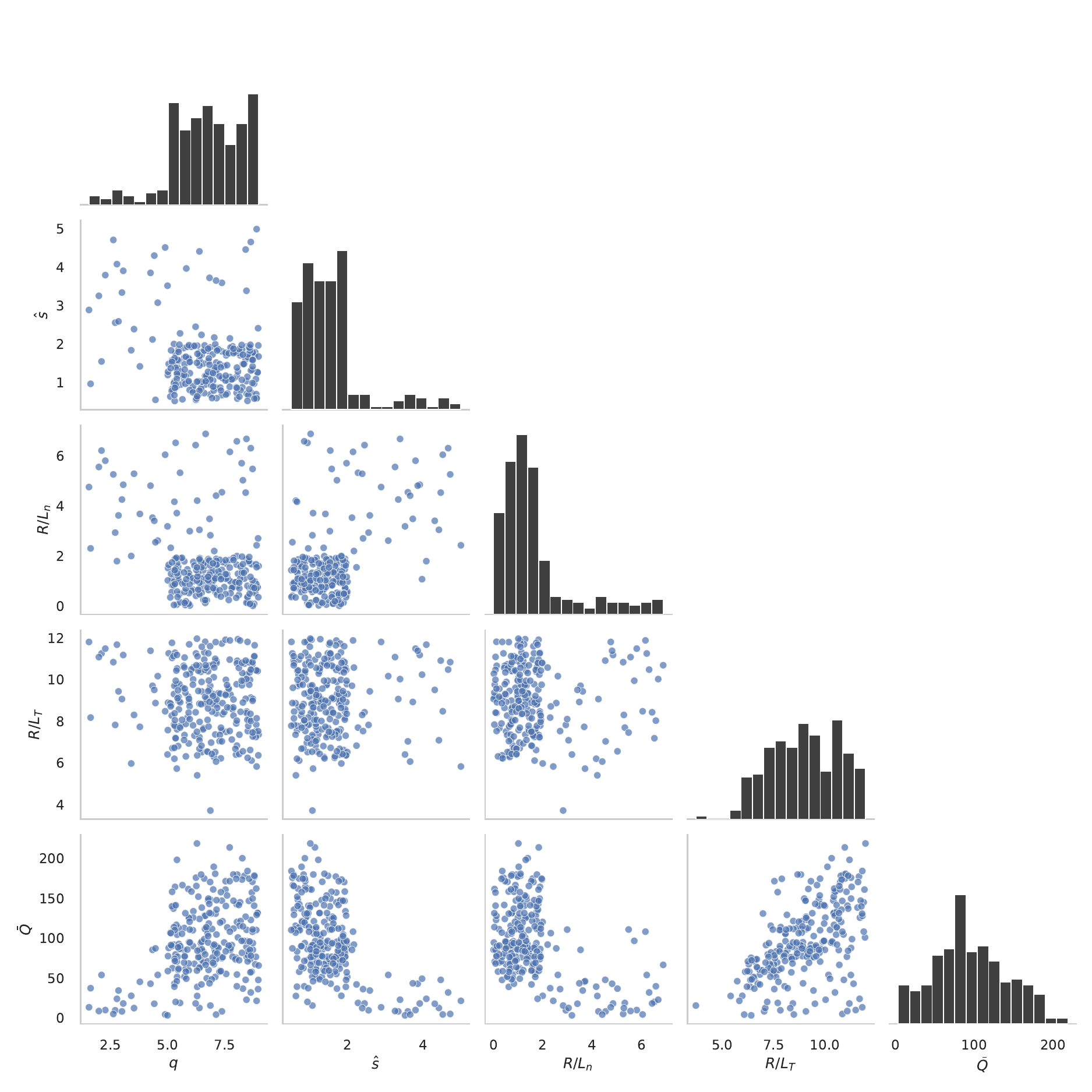}
    \caption{Distribution of input parameters $\hat{s}$, $q$, $R/L_n$, and $R/L_t$ along with average heat flux $\bar{Q}$. The sampled parameter space is evenly distributed.}
    \label{fig:param_dist}
\end{figure}

In total we perform two data generation passes.
For the first pass, we specify the ranges to sample the four parameters as $R/L_T \in [3,12]$, $R/L_n \in [1,7]$, $q \in [1,9]$, and $\hat{s} \in [0.5,5]$.
In addition, we also always vary the noise amplitude of the initial condition within $[1e-5, 1e-3]$ and the initial condition itself as one of $[\operatorname{sin},\operatorname{cos}, \operatorname{random} ]$.
We generate 100 simulations in total with this setup, but found that only approximately half of them actually exhibit turbulence.
The turbulent configurations are depicted as the sparse point cloud in \cref{fig:param_dist}.
Therefore, we adapt the parameter ranges for the second data generation pass to $R/L_T \in [6,12]$, $R/L_n \in [0,2]$, $q \in [5,9]$, and $\hat{s} \in [0.5,2]$.
Effectively, this change in parameters reduces stabilizing factors such as magnetic shearing and density gradients to force simulations to exhibit turbulence.
We sample another 200 simulations from these parameter ranges and found that all of these develop strong turbulence (see distribution of $\bar{Q}$ in \cref{fig:param_dist}).
Eventually, we end up with around 250 simulations of which we randomly select three as a validation set during training and  six as in-distribution evaluation set, to end up with a final training set of 241 simulations.
Importantly, we still include the initially generated turbulent simulations as data generation is computationally expensive.
Finally, we use the same parameter combination of each nonlinear simulation to generate their linear counterparts for the reduced-order model.

\begin{figure}
    \centering
    \includegraphics[width=\linewidth]{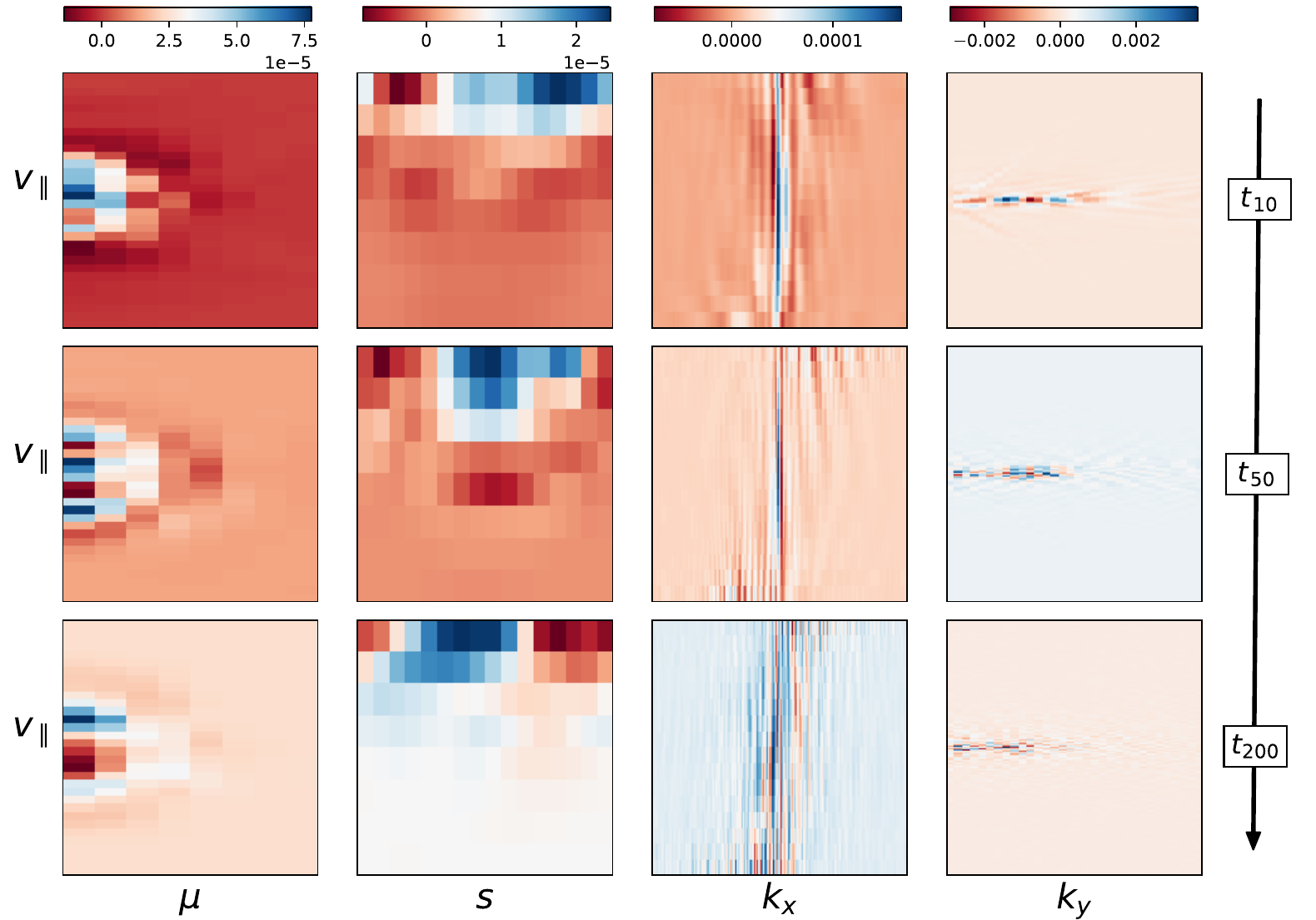}
    \caption{Visualization of the 5D distribution function of nonlinear gyrokinetics (ground truth, Fourier space along $k_x$ and $k_y$). We show different combinations of axes of the 5D field while averaging over the remaining ones for different timesteps. Colorbars are shared columnwise.}
    \label{fig:dft_vis}
\end{figure}

\textbfp{5D Data Visualization}
We show a visual illustration of the 5D distribution function in \cref{fig:dft_vis}.
For visualization purposes, we always show combinations of the different axes while averaging or slicing across the remaining ones.
This way, we end up with 2D planes that are easy to visualize.

\begin{figure}
    \centering
    \includegraphics[width=\linewidth]{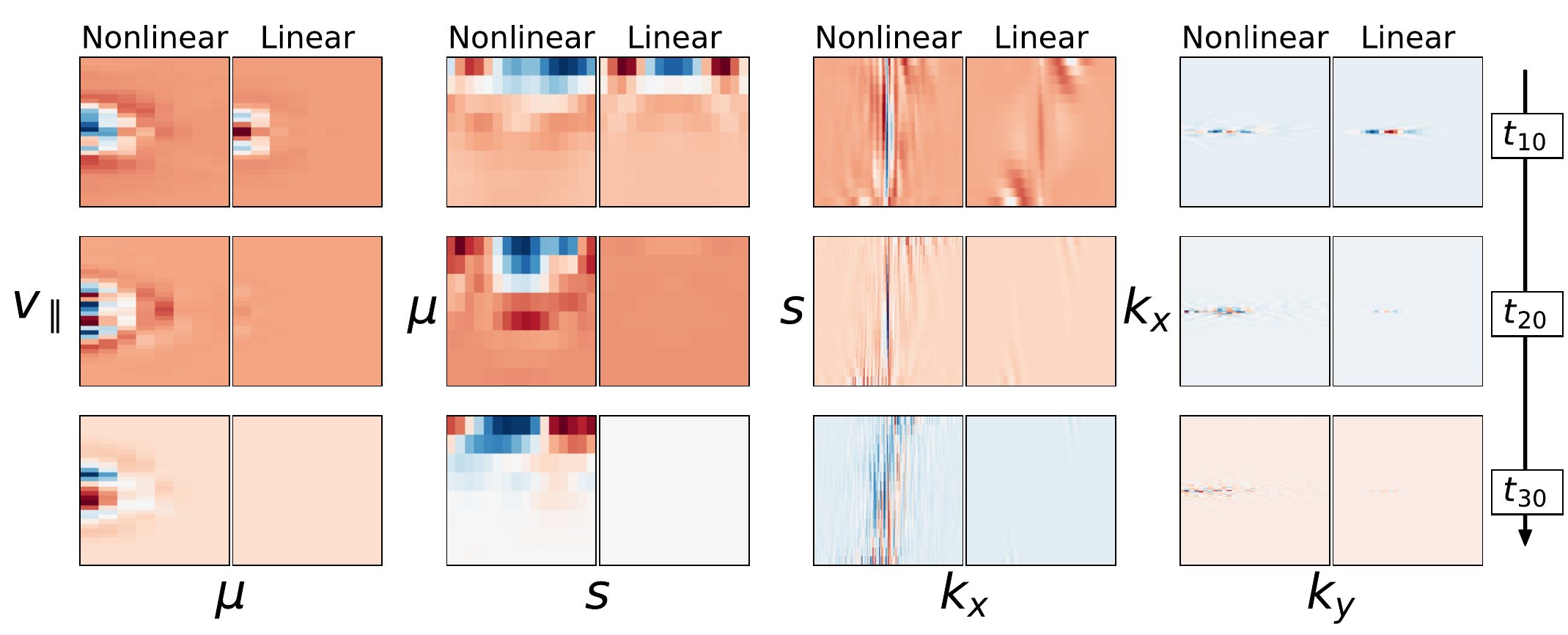}
    \caption{Comparison of 5D distribution functions for linear and nonlinear simulations (ground truth, Fourier space along $k_x$ and $k_y$). We show different combinations of axes of the 5D field while averaging over the remaining ones for different timesteps. Colorbars are shared across each 2D nonlinear and linear plane.}
    \label{fig:dft_nonlin_lin_vis}
\end{figure}

Furthermore, we also highlight the difference between nonlinear and linear simulations in \cref{fig:dft_nonlin_lin_vis}.
We plot 2D planes of the 5D field of nonlinear and linear simulations side-by-side for three different timesteps.
Since colorbars are shared across each pair of 2D planes, the difference between nonlinear and linear simulations can be clearly observed.
Specifically, the structure in the linear simulation is symmetric, while the nonlinear simulation exhibits asymmetries.
Furthermore, the difference is particularly obvious in wavenumber space, where (1) magnitudes of nonlinear simulations are much larger, and (2), there is a lot more interaction between modes.
Hence, there is structure in the nonlinear simulations that is not captured in linear ones, which is a drawback of approximations based on linear simulations, i.e. for quasilinear models.

\section{Implementation Details}
\label{app:impl_details}

\paragraph{Data.}
All trajectories are nonlinear, electrostatic ($\beta=0$), collisionless flux-tube
simulations with adiabatic electrons generated with GKW around the Cyclone Base Case.
The perturbed ion distribution function $\delta f$ is stored on a grid of
$32\,(v_\parallel)\times 8\,(\mu)\times 16\,(s)\times 85\,(x)\times 32\,(k_y)$,
with the radial direction transformed to real space and the binormal direction kept in
Fourier space (real and imaginary parts as channels). 
The zonal ($k_y=0$) component is split into two additional channels, giving 4 input/output channels (channelwise mode separation).
The electrostatic potential $\phi$ is stored in real space on the $16\times 85\times 32$
$(s,x,y)$ grid with a single channel.
Inputs and targets are $z$-score normalised with dataset-level statistics; $\delta f$
statistics are computed separately for each $\mu$ index, $\phi$ statistics are global.
We standardize all fields and fluxes to zero mean and unit variance.
To enable autoregressive rollouts, we simply accumulate all statistics across all training simulations, and use them to normalize and denormalize during inference.

\paragraph{Task.}
GyroSwin performs one-step prediction
$\delta f(t)\mapsto\big(\delta f(t+\Delta t),\,\phi(t+\Delta t)\big)$ conditioned on
$(t, R/L_T, R/L_n, \hat s, q)$. 
During inference we rollout the model to match the original number of timesteps of the original simulation.
The 241 training trajectories result in 44{,}400 one-step training pairs per epoch.

\paragraph{Architecture.}
GyroSwin is a two-branch shifted-window (Swin) transformer U-Net.
The $\delta f$ branch operates on a 4D token grid.
The $\mu$ axis is folded into the
channel dimension (after adding a learned positional embedding over $\mu$), so windowed
attention runs over $(v_\parallel, s, x, k_y)$.
The $\phi$ branch is a 3D Swin decoder over $(s,x,y)$ that has no encoder of its own.
Its latents are obtained from the $\delta f$ latents through attention-based velocity-space integration (a learned query attends over all $v_\parallel$ tokens at each spatial location, 8 heads) at every resolution level.
Both branches exchange information through bidirectional cross-attention mixing blocks (8 heads, MLP ratio 2, attention dropout 0.1) at the bottleneck, in every decoder stage, and before unpatching.
Components shared by all model sizes:
\begin{itemize}\setlength\itemsep{0pt}
  \item \textbf{Patching:} patch size $(4,2,4,5,4)$ over $(v_\parallel,\mu,s,x,k_y)$,
        giving an $8\times4\times17\times8$ token grid (4{,}352 tokens) for $\delta f$ and a
        $4\times17\times8$ grid for $\phi$ (patch size $(4,5,4)$).
        Patch embedding is a 2-layer MLP (hidden ratio 4, GELU \citep{hendrycks2016gaussian}); unpatching is a 2-layer MLP
        (hidden ratio 8, LeakyReLU) applied to the concatenation of decoder output and the
        patch-embedded input (patch skip).
  \item \textbf{Windows:} window size $(4,2,9,4)$ on the $\delta f$ grid and $(2,9,4)$ on the
        $\phi$ grid, with shifted windows in alternating blocks.
  \item \textbf{U-Net depth:} one down/up level. Encoder stage at width $d$ $\to$ patch
        merging ($2\times$ downsampling on all axes with more than two tokens, width $2d$)
        $\to$ bottleneck of 2 windowed Swin blocks with 8 heads $\to$ patch expansion $\to$
        decoder stage at width $d$ with skip connection.
  \item \textbf{Swin block:} pre-norm LayerNorm without affine parameters, windowed
        multi-head self-attention with SwinV2 log-spaced continuous relative position bias
        (2-layer MLP, 512 hidden units), no absolute or rotary position embeddings, MLP with
        hidden ratio 8 and GELU, stochastic depth rate 0.1.
  \item \textbf{Conditioning:} the five scalar conditions are embedded with sinusoidal
        features (128 dims, max.\ wavelength $10^4$) followed by a linear layer and SiLU
        (512 dims). The embedding modulates every Swin block and the unpatching MLP through FiLM \citep{perez_film_2018} (per-channel scale and shift). We also experimented with DiT-style conditioning \citep{peebles_dit_2023}, but found no improvements over FiLM.
  \item \textbf{Initialisation:} Kaiming-uniform \citep{he_delving_2015} for attention and MLP weights and for the
        patching MLPs; small-variance normal for conditioning layers.
\end{itemize}

\paragraph{Loss.}
For both fields we use the relative squared error per sample,
$\mathcal L_u=\mathbb E\big[\|\hat u-u\|_2^2/\|u\|_2^2\big]$, computed on normalised data.
The total loss is $\mathcal L=\mathcal L_{\delta f}+w_\phi(\tau)\,\mathcal L_\phi$, where
$\tau$ is the training progress and $w_\phi$ is $0$ for $\tau<0.5$, increases linearly to
$1$ between $\tau=0.5$ and $\tau=0.7$, and stays at $1$ afterwards.

\paragraph{Optimisation.}
Hyperparameters related to optimisation are shown in \cref{tab:hyperparams}.

\begin{table}[h]
\centering\small
\caption{Hyperparameters used during optimisation.}
\begin{tabular}{ll}
\toprule
Optimiser & Adam \citep{kingma_adam_2015}, $\beta=(0.9,\,0.95)$ \\
Peak learning rate & $3\times10^{-4}$ (identical for all model sizes) \\
Schedule & linear warm-up over the first $1/6$ of all steps, cosine decay to $0$ \\
Weight decay & $10^{-5}$ ($L_2$, coupled; not applied to conditioning parameters) \\
Gradient clipping & global norm $1.0$ \\
Precision & bf16 autocast, fp32 master weights \\
Batch size & 8 per GPU, 1{,}024 global \\
Epochs & 500 ($\approx 43$ updates per epoch, $\approx 21.7$k updates in total) \\
Hardware & 128 NVIDIA GH200 GPUs (32 nodes $\times$ 4), PyTorch DDP \\
Validation & every 20 epochs, 2-step autoregressive rollout on 3 held-out trajectories \\
Model selection & lowest validation error on $\delta f$ \\
\bottomrule
\end{tabular}
\label{tab:hyperparams}
\end{table}

\paragraph{Heat-flux fine-tuning (second stage).}
To predict the time-averaged ion heat flux we attach a flux head to the frozen pretrained
backbone. 
The head applies a one-layer FiLM-conditioned cross-attention transformer (8 heads,
dropout 0.1) to the paired $(\phi,\delta f)$ latents at every resolution level, max-pools the
result over tokens, concatenates the pooled vectors across levels, and maps them to a scalar with a 2-layer MLP. 
The regression target is the heat flux averaged over the last 80 snapshots of the trajectory. Only the flux head is trained, with an $L_1$ loss, for 200 epochs using the optimiser settings above.
The checkpoint with the lowest validation flux MSE is kept.

\paragraph{Baselines}
We elaborate on implementation details of the reported baselines below.

\textbfp{Gaussian Process Regression (GPR)}
We use a four-parameter GPR trained on the operating parameters ($R/L_T$, $R/L_n$, $q$, and $\hat{s}$). It uses a Matern $3/2$ kernel \citep{rasmussen_kernels_2006} to predict the nonlinear flux. This is the model proposed by \citet{Hornsby2024}.

\textbfp{Multi-Layer Perceptron (MLP)}
We use a 3 layer MLP with 128 hidden dimension and GELU activations. The inputs are the four operating parameters ($R/L_T$, $R/L_n$, $q$, and $\hat{s}$), which are embedded in a continuous sin-cos space, and the output is the time-averaged scalar heat flux. This is similar to QLKNN proposed by \citet{plassche_qlknn_2020}, but trained on nonlinear fluxes instead of quasiliner ones.

\textbfp{FNO}
The \texttt{neuraloperator} library\footnote{\url{https://github.com/neuraloperator/neuraloperator}} \citep{kovachki_neural_2023} contains tensorized nD implementations of Fourier convolution layers. However, as shown in \cref{tab:main_diagnostics} these are too expensive to be trained on our data.
To enable training FNO \citep{li_neural_2020}, we utilize a much faster 3D FNO which operates only on the spatial dimensions, while flattening the velocity space in the channels.
This baseline is shown in \cref{tab:main_results} and has 256 latent dimension, 4 layers and considers $\frac{1}{2}$ of the total modes on each dimension (5,004 flat spatial modes).

\textbfp{PointNet}
We use PointNet's \citep{qi2016pointnet} implementation from SIMSHIFT \citep{setinek2025simshiftbenchmarkadaptingneural}, adapted to work on regular grid data.
The total number of points on the grid is approximately 22.3 million considering both channel dimensions.
Therefore, to train this coordinate-based baseline, we randomly subsample the grid to 65,536 points for each sample.
We also experimented with larger number of coordinates per-batch but observed diminishing returns beyond 65,536 points.
Since the task is to predict the evolution of the distribution function, we provide its values at the corresponding coordinates at time $t$ to predict those for $t+1$.
For embedding grid coordinates we employ per-dimension embedding layers of the corresponding resolution and concatenate them to obtain the final coordinate embedding.
Each of these embeddings is of dimension 64 leading to a hidden dimension of 320 for 5D input.
We use GELU activations and condition the concatenated values and coordinates on the four operating parameters.
The encoder projects the conditioned coordinates/values to dimensionality 640.
Finally, we project to 12,800 global pooling features which are concatenated with the local coordinate features and serve as input to the decoder that predicts the field values at the local coordinates for the next timestep.

\textbfp{Transolver}
We use Transover's \citep{wu2024Transolver} implementation from SIMSHIFT \citep{setinek2025simshiftbenchmarkadaptingneural}, adapted to work on regular grid data.
We follow the same sampling and coordinate strategy as for PointNet.
The Transolver embedding dimension is set to 512 with 8 transformer blocks using 8 attention heads and 256 slices per head.
Again, we provide the values of the distribution function at time $t$ as input to predict the values at $t+1$.

\section{Scaling Configurations}
\label{app:scaling_configs}

We scale \ourmethod{} along three axes while keeping every hyperparameter of
Appendix~\ref{app:impl_details} fixed. 

\paragraph{Model axis.}
\cref{tab:gyroswin_sizes} lists the four model sizes. The width $d$ is the token
dimension in the encoder and decoder stages.
The bottleneck runs at $2d$ with 2 blocks and
8 heads for all sizes. 
Parameter counts are for the pretrained model (no flux head).
Forward FLOPs are measured per training sample and training compute is estimated as
$C\approx 3\times\mathrm{FLOPs}_{\mathrm{fwd}}\times\#\mathrm{samples}\times\#\mathrm{epochs}$.
Every model-axis point is trained with two seeds.

\begin{table}[h]
\centering\small
\caption{GyroSwin model sizes. Blocks per stage refers to the number of Swin blocks in the
encoder stage and in the decoder stage, respectively.}
\label{tab:gyroswin_sizes}
\begin{tabular}{lrrrrrr}
\toprule
Model & Width $d$ & Heads & Head dim & Blocks/stage & Params &  FLOPs$_{\mathrm{fwd}}$ \\
\midrule
XS     & 128  & 8  & 16 & 2 & 45.5M  & 0.31 TFLOP \\
Small  & 256  & 8  & 32 & 2 & 87.3M  & 0.44 TFLOP \\
Medium & 512  & 32 & 16 & 2 & 239.2M & 0.84 TFLOP \\
Large  & 1024 & 64 & 16 & 4 & 893.4M & 3.01 TFLOP \\
\bottomrule
\end{tabular}
\end{table}

\paragraph{Data axis.}
At fixed model size we train on nested subsets of the training trajectories (\cref{tab:gyroswin_data_sims}). 
Because the number of epochs is fixed, smaller subsets receive proportionally fewer parameter updates. 
The Medium model is trained with two seeds on every subset. 
The XS model with one seed.

\begin{table}[h]
\centering\small
\caption{Simulation-count subsets. Samples are one-step training pairs per epoch.}
\label{tab:gyroswin_data_sims}
\begin{tabular}{lrrl}
\toprule
Subset & Trajectories & Samples/epoch & Trained Models \\
\midrule
XS   & 37  & 6{,}845  & Medium, XS \\
S    & 63  & 11{,}655 & Medium, XS \\
M    & 150 & 27{,}380 & Medium, XS \\
ML   & 199 & 35{,}705 & Medium, XS \\
Full & 241 & 44{,}400 & all sizes \\
\bottomrule
\end{tabular}
\end{table}

\section{Additional Results}
\label{app:additional_results}

We include qualitative results for autoregressive rollouts of \ourmethod.
Specifically, we visualize model predictions for $f$ for the test case {$\mathbf{\text{ID}}_1$ and compare it to the ground truth in \cref{fig:dft_model_vs_gt} for timesteps $\{10, 100, 200\}$.
We observe that the overall structure is more or less preserved within the first 10 rollout steps, however, afterwards the model suffers from error accumulation.
This is especially pronounced for the velocity space, where the model seems to overpredict.
Interestingly, though, the predictions in wavenumber space remain somewhat stable and do not diverge, evan after 100 autoregressive prediction steps.
Generally we can say that \ourmethod{} yields stable predictions for long rollouts well beyond 100 timesteps.

\begin{figure}[bt]
    \centering
    \includegraphics[width=0.9\linewidth]{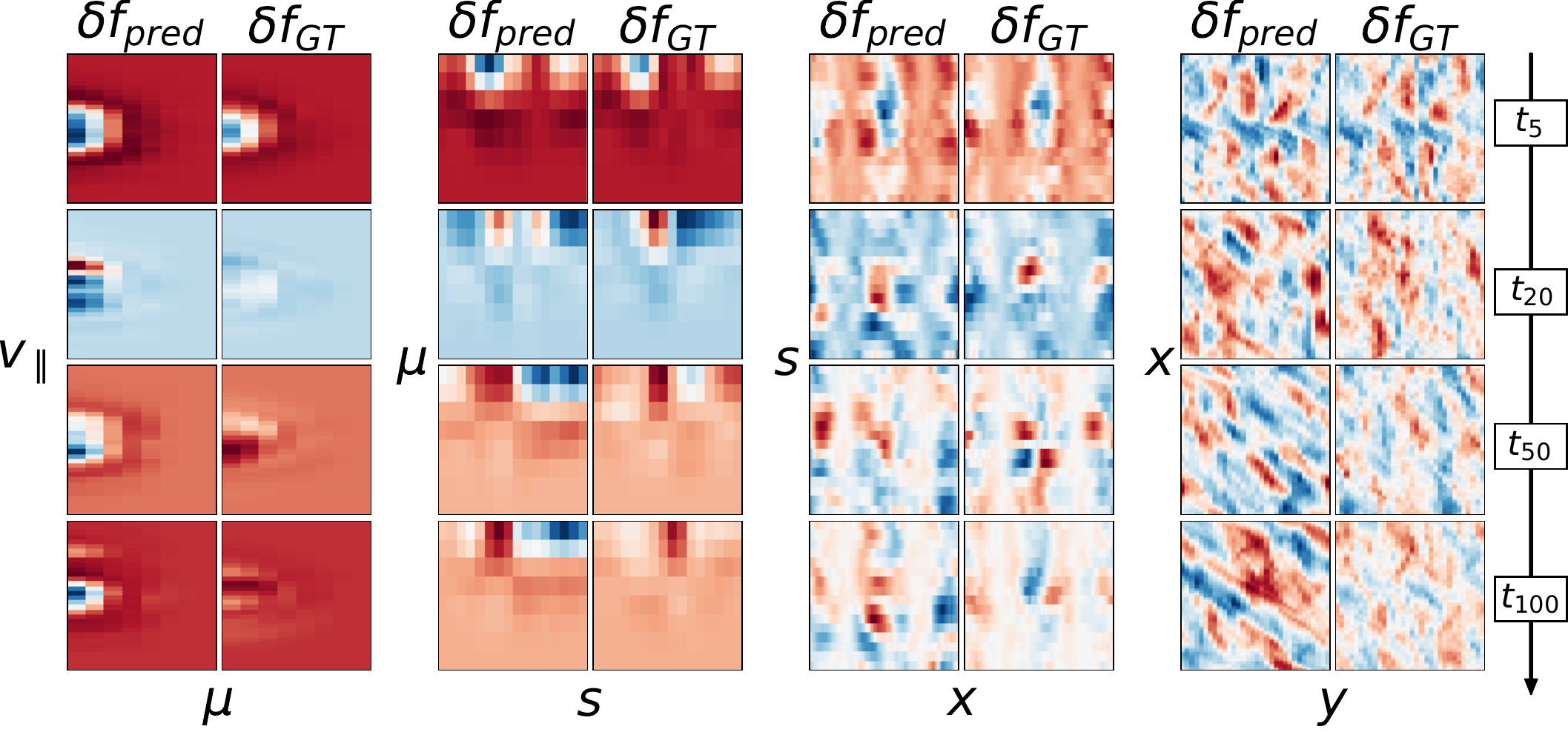}
    \caption{Side-by-side comparison of autoregressive rollout predictions with \ourmethod{} compared to ground truth ($x$ and $y$ in real space), in the saturated phase with shared colorbars. \ourmethod{} preserves the high-level structure within the first rollout steps. After a larger amount of rollout steps error accumulates, but predictions remain stable and do not diverge. }
    \label{fig:dft_model_vs_gt}
\end{figure}

\section{Diagnostics}
\label{app:diagnostics}
\begin{figure}[t]
    \centering
    \begin{subfigure}[b]{0.48\linewidth}
        \centering
        \includegraphics[width=\linewidth]{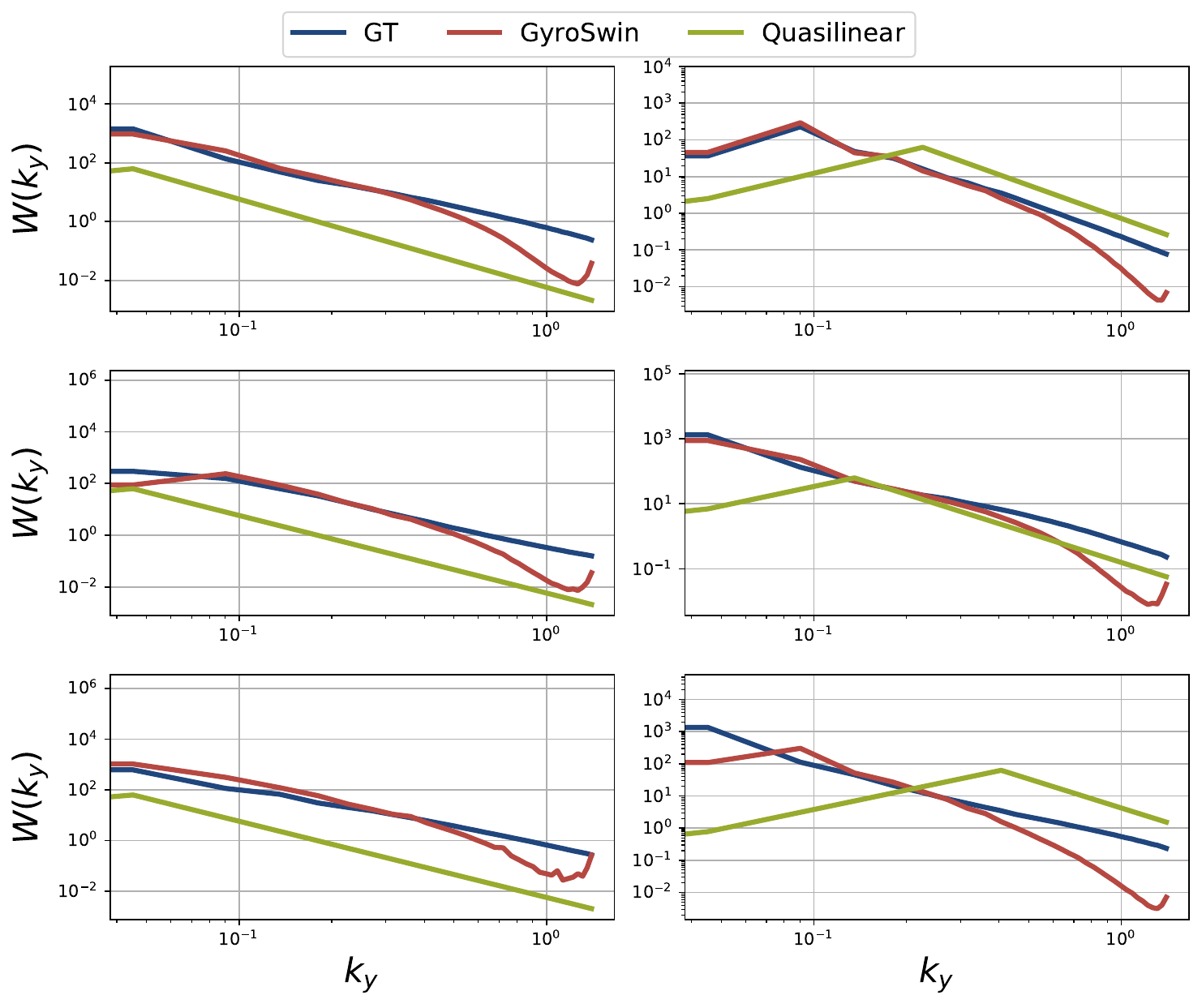}
        \caption{\textbf{ID}}
        \label{fig:id_kyspecs}
    \end{subfigure}
    \hfill
    \begin{subfigure}[b]{0.48\linewidth}
        \centering
        \includegraphics[width=\linewidth]{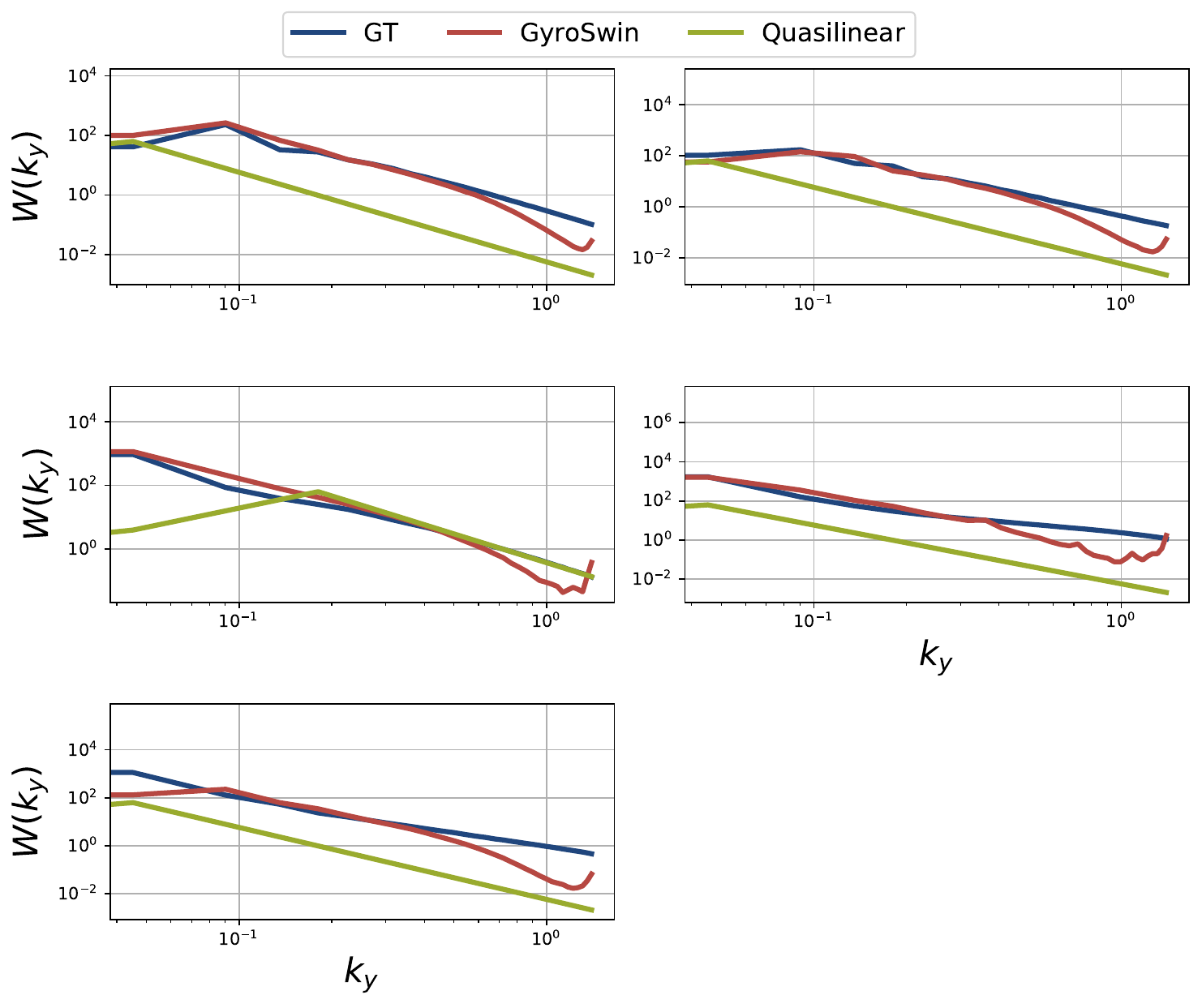}
        \caption{\textbf{OOD}}
        \label{fig:ood_kyspecs}
    \end{subfigure}
    \caption{Comparison of $W(k_y)$ In-Distribution and Out-Of-Distribution.}
    \label{fig:all_kyspecs}
\end{figure}

\begin{figure}[t]
    \centering
    \begin{subfigure}[b]{0.48\linewidth}
        \centering
        \includegraphics[width=\linewidth]{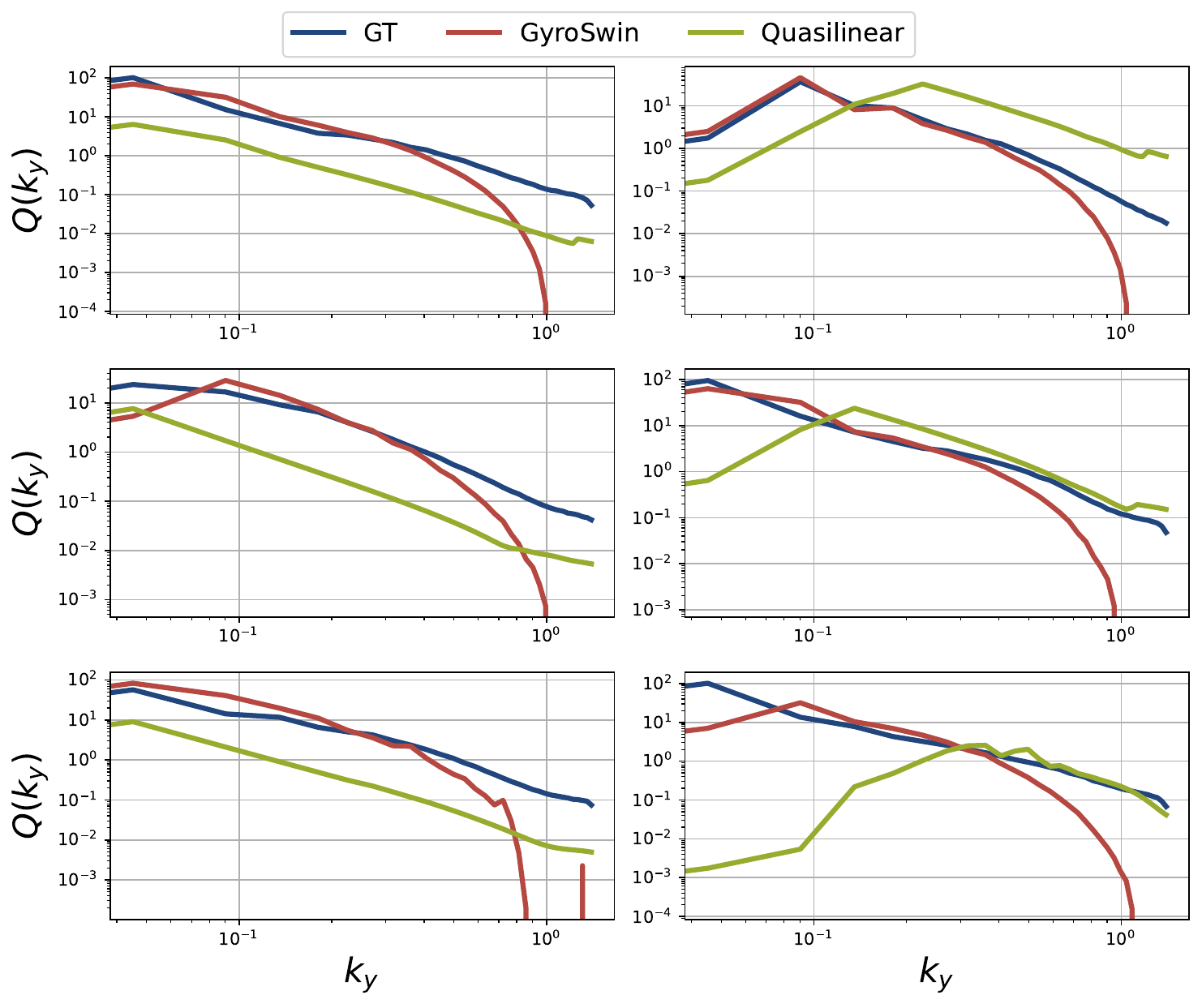}
        \caption{\textbf{ID}}
        \label{fig:id_fluxspecs}
    \end{subfigure}
    \hfill
    \begin{subfigure}[b]{0.48\linewidth}
        \centering
        \includegraphics[width=\linewidth]{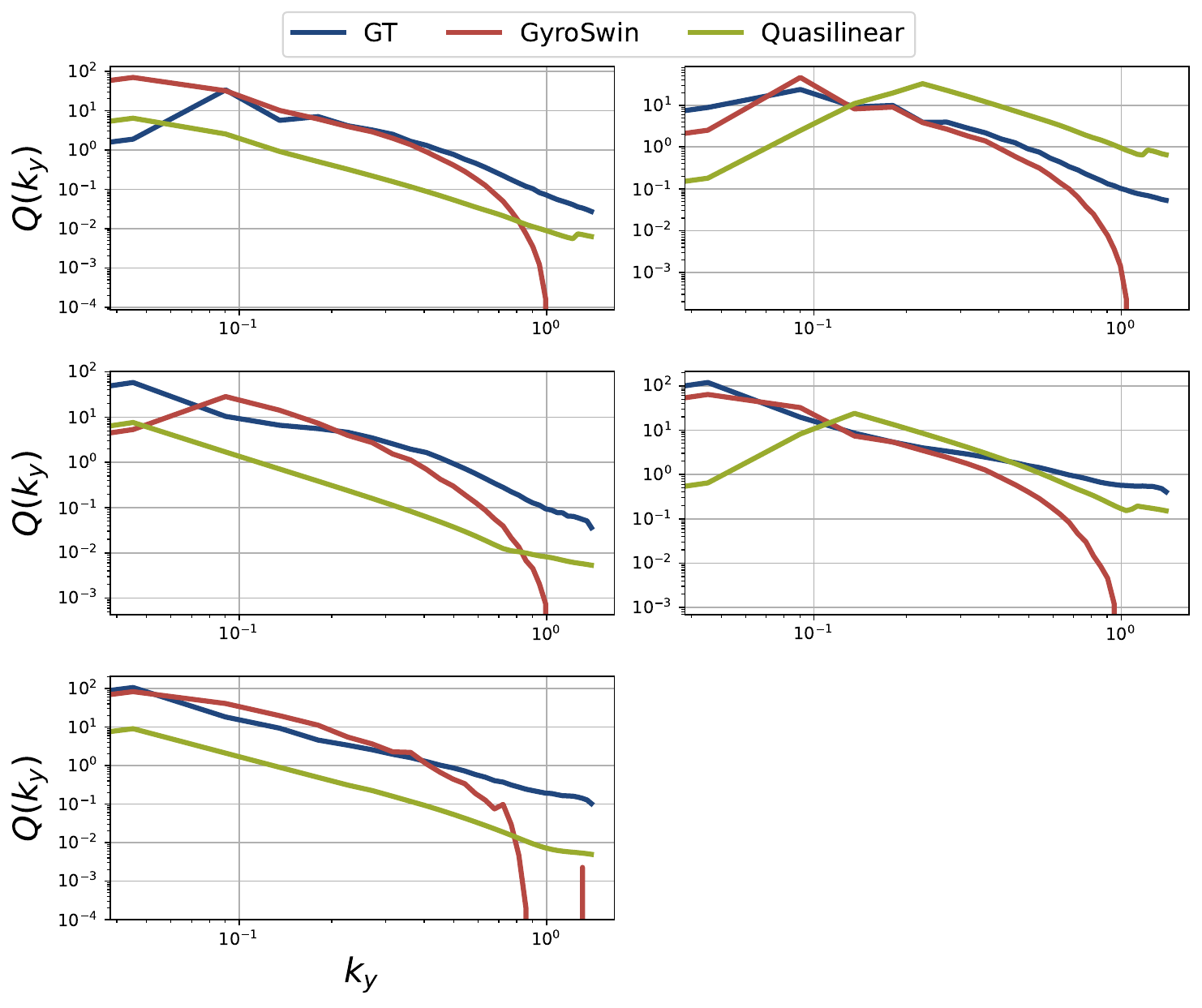}
        \caption{\textbf{OOD}}
        \label{fig:ood_fluxspecs}
    \end{subfigure}
    \caption{Comparison of flux spectra In-Distribution and Out-Of-Distribution.}
    \label{fig:all_fluxspecs}
\end{figure}

We provide visualizations for the time-averaged turbulence-intensity spectrum $W(k_y)$ for the six different \textbf{ID} test cases in \cref{fig:id_kyspecs} and the five different \textbf{OOD} test cases in \cref{fig:ood_kyspecs}. 
Furthermore, we show flux spectra $Q(k_y)$ for the six different \textbf{ID} test cases in \cref{fig:id_fluxspecs} and the five different \textbf{OOD} test cases in \cref{fig:ood_fluxspecs}. 
Generally, we observe that \ourmethod{} susbtantially improves over the state-of-the-art reduced-numerical quasilinear model on $W(ky)$, which is reflected in the pearson correlation coefficients shown in \cref{tab:main_diagnostics}.
Interestingly, the fit for $Q(k_y)$ at first glance appears to be improved as well, however the correlation is approximately the same as for the quasilinear model.
The reason for this is that high-frequencies are not well captured by \ourmethod{}.
We expect this to be a result of the inherent spectral bias of neural network architectures \citep{rahaman_spectral_2019}. 

In addition to the time-averaged spectra, we provide visualizations for the time-averaged zonal flow profile for all \textbf{ID} simulations as well as \textbf{OOD} simulations for \ourmethodlarge.
We observe that \ourmethodlarge usually tends to overestimate the amplitude therefore we normalize it for visualization purposes.
The results for the \textbf{ID} and \textbf{OOD} test cases can be observed in \cref{fig:id_zfprofiles}, and \cref{fig:ood_zfprofiles}, respectively.
Generally, we observe that the predicted zonal flow profile of \ourmethodlarge appears as a smoothed out version of the ground-truth.
Again, our intuition is that this is due to the inherent low-frequency bias.
Interestingly, on some test cases the zonal flow profile is entirely off, which explains the rather high variance observed in the main table.
Overall, we can conclude that \ourmethodlarge{} can capture the zonal flow of unseen simulations in an autoregressive manner even though there is plenty of room for improvement.

\begin{figure}[t]
    \centering
    \begin{subfigure}[b]{0.48\linewidth}
        \centering
        \includegraphics[width=\linewidth]{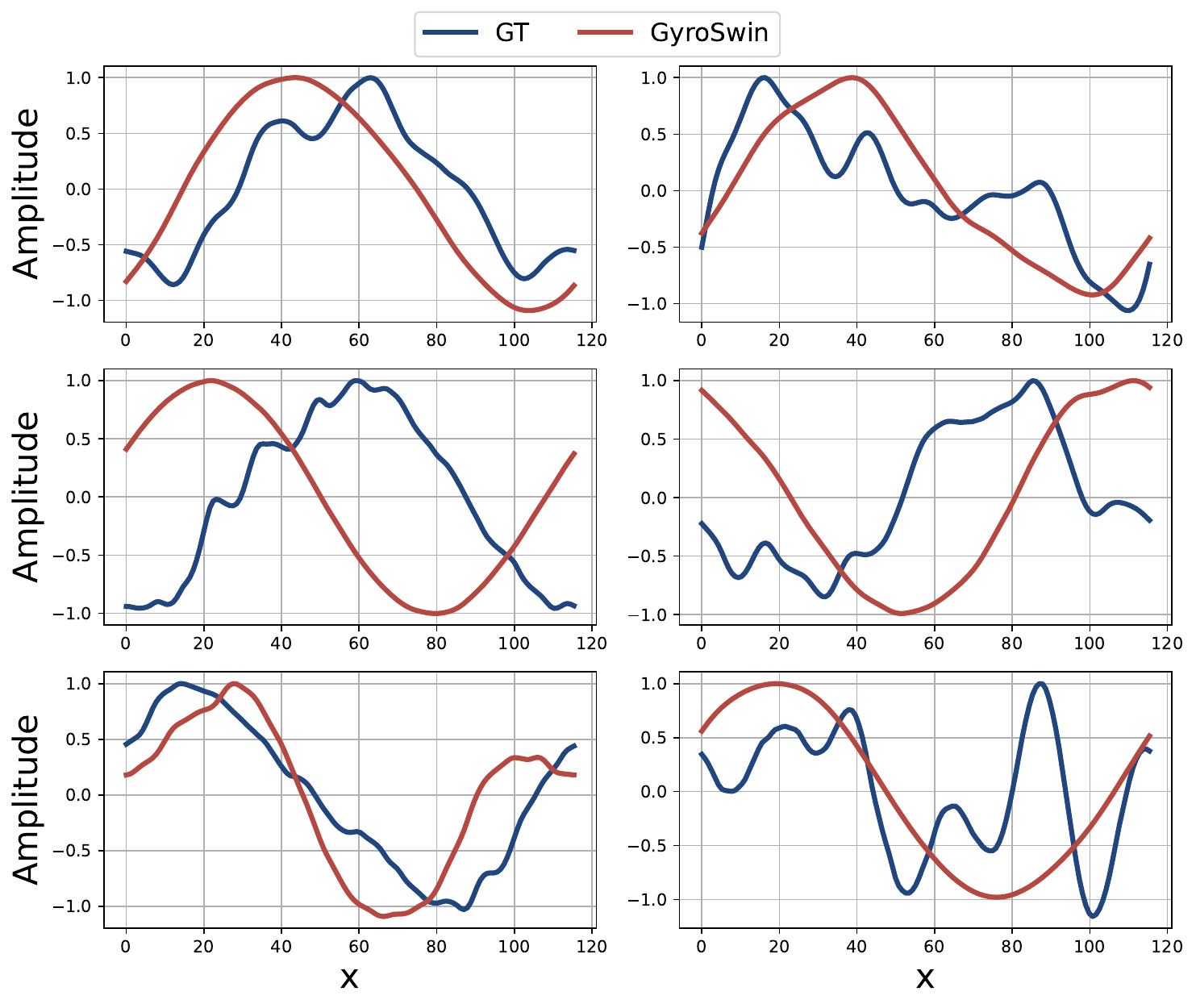}
        \caption{\textbf{ID}}
        \label{fig:id_zfprofiles}
    \end{subfigure}
    \hfill
    \begin{subfigure}[b]{0.48\linewidth}
        \centering
        \includegraphics[width=\linewidth]{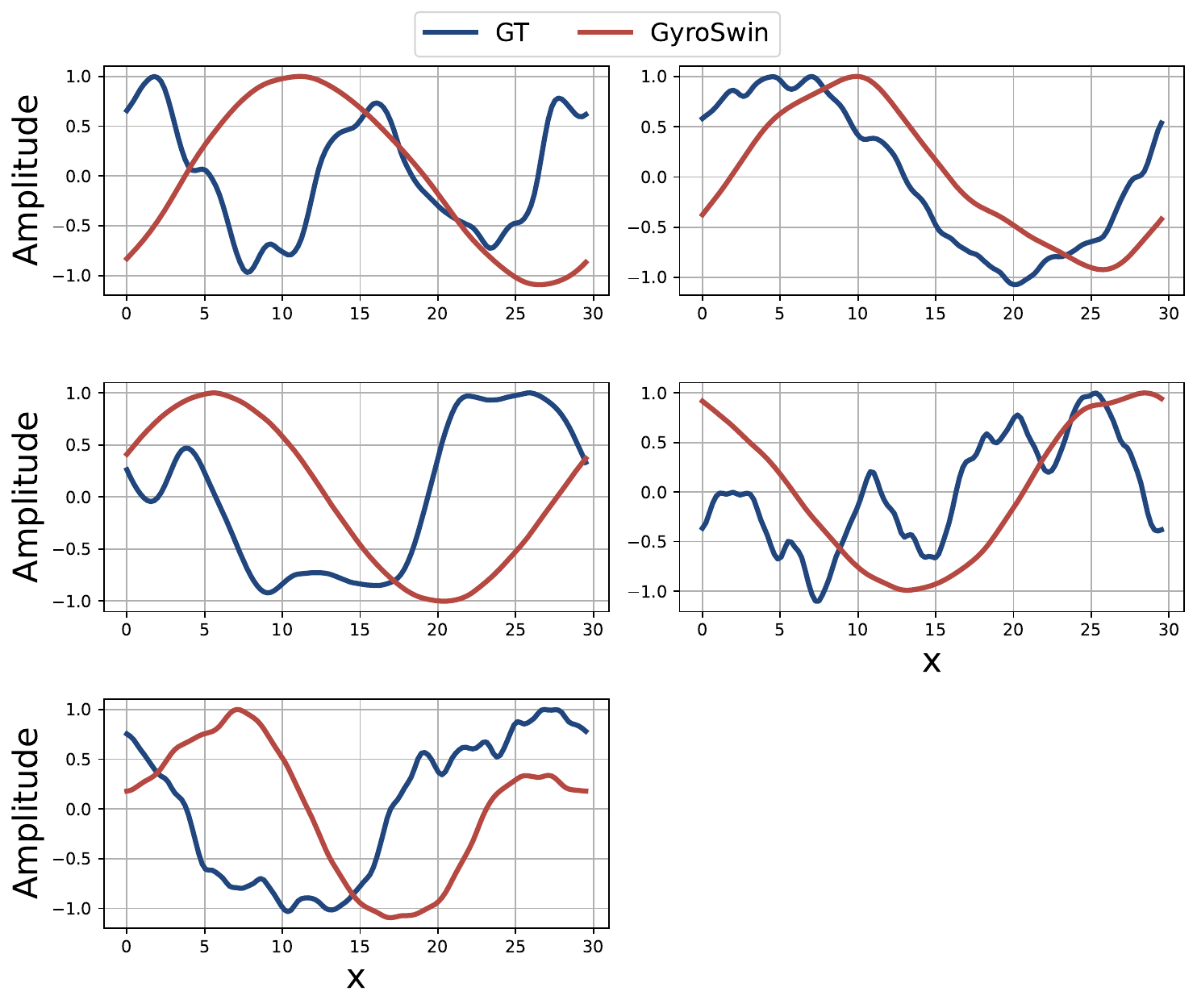}
        \caption{\textbf{OOD}}
        \label{fig:ood_zfprofiles}
    \end{subfigure}
    \caption{Comparison of zonal flow profiles In-Distribution and Out-Of-Distribution.}
    \label{fig:all_zfprofiles}
\end{figure}

\section{Ablation Studies}
\label{app:ablation_studies}

To justify the design choices of \ourmethod{}, we conduct a set of ablation studies as follows.
We progressively add components of \ourmethod{} to a 5D Swin transformer and evaluate based on correlation time to the 5D distribution function $f$ and RMSE for the time-averaged flux trace $\bar{Q}$.
We report our results in \cref{app:ablation_studies}.
Interestingly, the 5D-Swin already appears to be capable of performing stable rollouts, however it completely fails in capturing the nonlinear flux trace.
When moving to a UNet like structure, we observe a slight drop in correlation time, but improved flux predictions.
Furthermore, adding channelwise mode separation gives a boost in correlation time, particularly on \textbf{OOD} simulations.
Moreover, there is a slight improvement in correlation time when adding our latent cross-attention and integrator modules.
Finally, the most significant boost is observed when adding the flux prediction head, which results in our final design of \ourmethod{}, which achieves the best combination of stable autoregressive rollouts and nonlinear flux prediction.

\begin{table}[h]
\centering
\caption{Ablation study on different components in \ourmethod{}, i.e. channel-wise mode separation of zonal flows and latent cross-attention/integrator modules. We report correlation time with $\tau = 0.1$ for the 5D distribution function and RMSE for $\bar{Q}$ on \textbf{ID} and \textbf{OOD} test cases. All methods are trained on a reduced dataset of 48 simulations.}
\resizebox{\textwidth}{!}{
\begin{tabular}{l|c|cc|cc}
\toprule
\multirow{2}{*}{\textbf{Method}} & \multirow{2}{*}{\textbf{Params [M]}} & \multicolumn{2}{c|}{$f$} & \multicolumn{2}{c}{$\bar{Q}$}  \\
& & \textbf{ID} ($\uparrow$) & \textbf{OOD} ($\uparrow$) & \textbf{ID} ($\downarrow$) & \textbf{OOD} ($\downarrow$) \\
\midrule
5D-Swin & 46.2 & 24.83 $\pm$ 3.19 & 25.00 $\pm$ 2.39 & 121.34 $\pm$ 13.36 & 125.74 $\pm$ 23.23 \\
$\text{\ourmethod{}}_{f}$ & 36.3 & 19.50 $\pm$ 1.86 & 18.60 $\pm$ 0.51 & 119.85 $\pm$ 13.15 & 124.91 $\pm$ 23.2 \\
$\text{\ourmethod{}}_{f}$ + ZF channel & 44.2 & 19.33 $\pm$ 1.84 & 22.00 $\pm$ 2.26 & 119.19 $\pm$ 13.65 & 122.76 $\pm$ 23.63 \\
$\text{\ourmethod{}}_{f + \phi}$ & 77.8 & 20.17 $\pm$ 0.75 & 21.80 $\pm$ 1.20 & 106.26 $\pm$ 14.09 & 112.95 $\pm$ 25.22 \\
$\text{\ourmethod{}}_{f + \phi}$ + LatentCA & 87.3 & 21.00 $\pm$ 2.62 & 22.80 $\pm$ 2.89 & 116.49 $\pm$ 12.81 & 121.41 $\pm$ 23.45 \\
$\text{\ourmethod{}}_{f + \phi + \bar{Q}}$ & 91.1 & 17.67 $\pm$ 2.16 & 19.40 $\pm$ 1.46 & 47.35 $\pm$ 6.20	& 59.15 $\pm$ 15.62 \\
\bottomrule
\end{tabular}
}
\label{tab:ablation_studies}
\end{table}

\section{Learning a saturation rule}
\label{app:linear_ablation}

We perform an additional ablation study where we discard the nonlinear term in \cref{gyrovlas}, e.g. train \ourmethod{} on linear simulations to predict nonlinear fluxes.
This effectively mirrors a saturation in the Quasilinear setting, except that it does not rely on growth rates and reconstruction of the $k_y$ spectrum.
Since linear simulations essentially converge to a fixed structure in the statistically steady state, we only use the last snapshot of a simulation and provide it as input to the model to predict the average nonlinear flux $\bar{Q}$.
We call this variant \ourmethodlinear{} and compare it to the Quasilinear saturation rule as well as the tabular regressors.
Since this baseline is relatively cheap to train, we train it for three seeds and for both training sets, containing 48 and 241 simulations and report our results in \cref{saturation_rule}.

Interestingly, we observe a consistent trend that machine learning based methods outperform the reduced numerical model.
An interesting aspect of this result is that the quasilinear saturation rule leverages additonal data such as linear growth rates of modes and flux contributions per mode which are hidden to tabular regressors and the learned saturation rule.
Furthermore, with increasing training data \ourmethodlinear{} outperforms tabular regressors GPR and MLP that map from operating parameters to nonlinear heat flux.
Since \ourmethodlinear{} is only trained on the linear part of \cref{gyrovlas}, therefore we can conclude that it is capable of generalizing over linear mode structures to infer nonlinear fluxes.
Finally, all methods experience improvements for larger training sets.
For the QL model the improvement can be traced back to a larger scaling factor that is fitted on more simulations, whereas all other methods are machin-learning based, hence naturally improve with more available training data.

\begin{table}[h]
\centering
\caption{Learning a saturation rule with \ourmethod. We report RMSE for the average flux $\bar{Q}$ on \textbf{ID} and \textbf{OOD} test cases. \ourmethodlinear{} significantly outperforms the reduced numerical model (QL).}
\begin{tabular}{l|cc|cc}
\toprule
\multirow{2}{*}{\textbf{Method}} & \multicolumn{2}{c}{48 sims} & \multicolumn{2}{c}{241 sims}  \\
& \textbf{ID} ($\downarrow$) & \textbf{OOD} ($\downarrow$) & \textbf{ID} ($\downarrow$) & \textbf{OOD} ($\downarrow$) \\
\midrule
\multicolumn{5}{c}{\textit{Reduced numerical methods}} \\
\midrule
QL & 89.48 $\pm$ 11.77 & 95.18 $\pm$ 21.58 & 56.68 $\pm$ 14.09 & 56.06 $\pm$ 14.35 \\
\midrule
\multicolumn{5}{c}{\textit{Tabular regressors}} \\
\midrule
GPR & 43.82 $\pm$ 10.84 & 59.28 $\pm$ 17.55 & 36.95 $\pm$ 9.24 & 25.89 $\pm$ 8.78 \\
MLP & 50.90 $\pm$ 10.87 & 62.88 $\pm$ 18.58 & 49.54 $\pm$ 13.36 & 34.99 $\pm$ 12.80 \\
\midrule
\multicolumn{5}{c}{\textit{Learned saturation rule}} \\
\midrule
$\text{GyroSwin}_{\text{Linear}}$  & 76.74 $\pm$ 11.79 & 82.95 $\pm$ 19.58 & 42.85 $\pm$ 12.65 & 23.40 $\pm$ 5.89 \\
\bottomrule
\end{tabular}
\label{tab:saturation_rule}
\end{table}

\subsection{Vision Transformer Complexity}
\label{app:swin_complexity}
Computational complexity of 5D swin layers, with an input of resolution $x \times y \times z \times h \times k$, $d$ channels and window size $M$ (assumed square for simplicity) is 
\begin{align*}
    O(\texttt{MSA)} &= 4 \: (xyzhw)^2 \: d^2 + 2 \: (xyzhw)^2 \: d \\
    O(\texttt{W-MSA)} &= 4 \: xyzhw \: d^2 + 2 M^5 \: xyzhw \: d
\end{align*}
removing the squared dependency on the sequence length $xyzhw$, and replacing it with a much smaller window complexity of $M^5$.
This consideration is the motivation behind the original swin paper by \citep{liu_swin_2021}, as it makes attention on images close to linear in resolution.

\end{document}